\documentclass[showpacs,pra,twocolumn,superscriptaddress,amsmath,amssymb,amsfonts]{revtex4-1}

\usepackage{graphicx}
\usepackage{fbpack}
\usepackage{fbpack}
\usepackage{color}

\newcommand{\aBS}{a^\mrm{BS}}
\newcommand{\adbs}{\ad\,\!^\mrm{BS}}
\newcommand{\smbs}{\sm^\mrm{BS}}
\newcommand{\spbs}{\spl^\mrm{BS}}
\newcommand{\szbs}{\sz^\mrm{BS}}
\newcommand{\drbs}{\Delta_r^\mrm{BS}}
\newcommand{\dqbs}{\Delta_a^\mrm{BS}}

\newcommand{\eq}[1]{(\ref{#1})}
\newcommand{\Tr}{\mrm{Tr}}

\begin{document}

\title{Dissipation and Ultrastrong Coupling in Circuit QED}

\author{F\'elix Beaudoin}
\affiliation{D\'epartement de Physique, Universit\'e de Sherbrooke, Sherbrooke, Qu\'ebec, Canada J1K 2R1}
\author{Jay M. Gambetta}
\affiliation{IBM T. J. Watson Research Center, P. O. Box 218, Yorktown Heights, NY 10598, USA}
\author{A. Blais}
\affiliation{D\'epartement de Physique, Universit\'e de Sherbrooke, Sherbrooke, Qu\'ebec, Canada J1K 2R1}

\date{\today}

\begin{abstract}
Cavity and circuit QED study light-matter interaction at its most fundamental level. Yet, this interaction is most often neglected when considering the coupling of this system with an environment. In this paper, we show how this simplification, which leads to the standard quantum optics master equation, is at the root of unphysical effects. Including qubit relaxation and dephasing, and cavity relaxation, we derive a master equation that takes into account the qubit-resonator coupling. Special attention is given to the ultrastrong coupling regime, where the failure of the quantum optical master equation is manifest. In this situation, our model predicts an asymmetry in the vacuum Rabi splitting that could be used to probe dephasing noise at unexplored frequencies. We also show how fluctuations in the qubit frequency can cause sideband transitions, squeezing, and Casimir-like photon generation.
\end{abstract}


\maketitle

\section{Introduction}

Elementary quantum mechanics teaches that a closed physical system always evolves in a reversible manner. However, control and readout imply a coupling of the quantum system to the outside world, making it subject to relaxation and decoherence. These irreversible dynamics, well understood from a theoretical viewpoint, have also been experimentally tested. In cavity QED~\cite{haroche:2006a}, macroscopic superpositions of quantum states of light have been built and their destruction due to their interaction with a reservoir has been observed~\cite{PhysRevLett.77.4887}. Using repeated QND measurements, the birth and death of single photons in a cavity has been studied~\cite{gleyzes2007quantum,guerlin2007progressive}. Circuit QED, a solid-state realization of cavity QED, also offers a detailed understanding of relaxation and dephasing phenomena~\cite{blais2004cavity,wallraff2004strong}. Spontaneous emission of a qubit in a resonator has been characterized with respect to the influence of far off-resonant modes~\cite{PhysRevLett.101.080502}. Moreover, the impact of measurement on qubit dephasing processes is well understood~\cite{gambetta2006qubit,gambetta2008quantum}, for instance in the cases of dispersive~\cite{PhysRevLett.94.123602} and bifurcation~\cite{PhysRevLett.106.167002} read-out.

Though both circuit and cavity QED  allow to study dissipation, solid-state devices allow much stronger light-matter interaction rates. For example, current-current coupling of a flux qubit to a Josephson junction in a resonator can boost the strength up to the order of the resonator and qubit frequencies~\cite{bourassa2009ultrastrong}, breaking the rotating-wave approximation (RWA). This ultrastrong coupling regime has been achieved experimentally with coplanar waveguides \cite{niemczyk2010circuit} and lumped LC resonators \cite{PhysRevLett.105.237001}. In parallel to these experimental efforts, dynamics of pure states have been theoretically studied~\cite{PhysRevLett.105.263603, PhysRevLett.105.023601, PhysRevA.82.062320, PhysRevA.83.030301}. A rigorous model based on the Bloch-Redfield formalism which describes photon losses has also been proposed by Hausinger and Grifoni~\cite{hausinger2008dissipative}. Finally, a non-Markovian model of dissipation has been used to predict the emission spectrum of an atom-resonator system in which the ultrastrong coupling strength is modulated over time~\cite{PhysRevA.74.033811,PhysRevA.80.053810} and to study the sensitivity of the system to noise in the qubit frequency~\cite{2011arXiv1106.1159N}.  In this paper, we give a complete description of dissipation including qubit relaxation and pure dephasing in the ultrastrong coupling regime, focusing on the standard case where the baths can be treated as Markovian.

Qubit-resonator coupling is at the heart of the problem with dissipation in the ultrastrong coupling regime. When the coupling between these two subsystems is small, interactions with the environment are treated separately for the qubit and the oscillator~\cite{haroche:2006a}. However, when the atom-field interaction increases up to the breakdown of the RWA, this approach leads to unphysical predictions. For example, and as will be illustrated later in Fig.~\ref{fig-relax}, relaxation baths bring the system out of its ground state even at $T=0$. Furthermore, in the presence of a strong qubit-resonator coupling, transitions at widely separated frequencies appear, breaking down the standard white noise approximation. To avoid such annoyances, the qubit-resonator coupling and colored baths must be included in the treatment of dissipation.

The outline of this work is as follows. First, we present the system Hamiltonian, along with a perturbative approach that can diagonalize it approximately. Next, we discuss the treatment of dissipation. In Section~\ref{sec-std}, we explain the standard approach to describe dissipation in the Jaynes-Cummings regime. We then show issues arising from the use of this technique in the ultrastrong regime, and devote Section~\ref{sec-model} to the presentation of a Lindbladian master equation that solves them. In Section~\ref{sec-strong-coupling}, we describe the implication of these results in the strong coupling regime. Finally, in Section~\ref{sec-ultrastrong}, we study physical consequences of the model obtained here in the Bloch-Siegert regime, for which counter-rotating terms can be treated in a perturbative fashion. We first show how the vacuum Rabi splitting spectrum is affected by non-RWA terms in the Hamiltonian and by the shape of the noise spectrum. We also introduce a potential technique to exploit these effects to study noise. We then present how the master equation helps to understand an analog of the time-dependent Casimir effect coming from pure qubit dephasing \cite{PhysRevA.78.053805}.

\section{Hamiltonians describing the qubit-resonator system \label{sec-hamiltonians}}

The Rabi Hamiltonian, describing the interaction of a two-level atom with a single electromagnetic mode of a resonator, takes the form ($\hbar=1$)~\cite{scully-zubairy}
\be	\label{eqn-Rabi}
	H_R=\omega_r\ad a + \frac{\omega_a}{2}\sz + g X\sx,
\ee
where $\omega_a$ is the qubit splitting, $\omega_r$ the resonator frequency, $g$ the coupling strength, and  $X=\ad+a$. In most experimental situations, $g \ll \omega_a, \omega_r$ and the rotating-wave approximation (RWA) can safely be made. This amounts to dropping the fast-oscillating, or counter-rotating, terms $I_\mrm{CR}=a\sm+\ad\spl$ from $H_R$. This approximation leads to the Jaynes-Cummings Hamiltonian~\cite{walls-milburn}
\be	\label{eqn-JC}
	H_{JC}=\omega_r\ad a + \frac{\omega_a}{2}\sz + g(a\spl+\ad\sm).
\ee
In opposition to the Rabi Hamiltonian, here the total number of quanta $N_q=(1+\sz)/2+\ad a$ is a good quantum number, allowing exact diagonalization of $H_{JC}$. The system enters the ultrastrong coupling regime when $g$ is so large with respect to $\omega_a$, $ \omega_r$ that $I_{CR}$ leads to experimentally observable consequences and the RWA cannot be safely made~\cite{niemczyk2010circuit, PhysRevLett.105.237001}. In this situation, since $\commut{N_q}{I_\mrm{CR}}\neq0$, the total number of excitations is not preserved, even though its parity is~\cite{PhysRevLett.105.263603}.  As a result, even in the ground state, the expected mean number of resonator and qubit excitations is non-zero.  

Although the analytical spectrum of $H_R$ has recently been found by Braak~\cite{PhysRevLett.107.100401}, it is defined in terms of the power series of a transcendental function. An approximate, but more simple form, can be found in the intermediate regime where g is small with respect to $\Sigma=\omega_a+\omega_r$, with the system still being in the ultrastrong coupling regime. This will be referred to as the Bloch-Siegert regime. This is  done using the unitary transformation~\cite{shavitt1980quasidegenerate,hausinger2008dissipative,PhysRevLett.105.237001}
\be	\label{eqn-unitary}
	U = \exp\left\{\Lambda(a\sm-\ad\spl)+\xi(a^2-a^\dagger\,\!^2)\sz\right\},
\ee
where $\Lambda=g/\Sigma$, and $\xi=g\Lambda/2\omega_r$ together with the Campbell-Baker-Hausdorff relation
\be\label{eqn-campbell}
	\eul{-X}H\eul{X}= H+\commut{H}{X}+\frac{1}{2!}\commut{\commut{H}{X}}{X}+\ldots
\ee
To second order in $\Lambda$, this yields the Bloch-Siegert Hamiltonian
\be
	U^\dagger H_R U \simeq H_\mrm{BS} = (\omega_r+\mu\sz)\ad a+\frac{\tilde{\omega}_q}{2}\sz+gI_+,
\ee
where $I_+=a\spl+\ad\sm$, $\tilde{\omega}_q=\omega_a+\mu$, and $\mu=g^2/\Sigma$. This Hamiltonian is similar to the Jaynes-Cummings Hamiltonian, but contains Bloch-Siegert shifts $\mu$ on qubit and resonator frequencies. 

Since $H_\mrm{BS}$ is block-diagonal, its eigenstates can be found exactly to be
\begin{align}
	\ket{n,+}&=-\sin\theta_n\ket{e,n-1}+\cos\theta_n\ket{g,n}\label{eqn-nplus}\\
	\ket{n,-}&=\;\;\;\cos\theta_n\ket{e,n-1}+\sin\theta_n\ket{g,n},\label{eqn-nminus}
\end{align}
with the Bloch-Siegert mixing angle
\be
	\theta_n=\arctan\left[\frac{\Delta_n^\mrm{BS}-\sqrt{(\Delta_n^\mrm{BS})^2+4g^2n}}{2g\sqrt{n}}\right],\label{eqn-angle}
\ee
and where $\Delta_n^\mrm{BS}=\omega_a-\omega_r+2\mu n$. To second order in $\Lambda = g/\Sigma$, the excited eigenstates $\ket{\widetilde{n,\pm}}$ of the Rabi Hamiltonian in the bare basis are then given by
\be\label{eq:BS_Basis_tranformation}
	\ket{\widetilde{n,\pm}}=U\ket{n,\pm},
\ee
while the ground state takes the form
\be
	\ket{\widetilde{g0}}=U\ket{g0}\simeq\left(1-\frac{\Lambda^2}{2}\right)\ket{g0}-\Lambda\ket{e1}+\xi\sqrt2\ket{g2}.
\ee
As mentioned before, the ground state is no longer the simple $H_{JC}$ ground state $\ket{g0}$, but now contains qubit-resonator excitations.

Unitary transformation Eq.~(\ref{eqn-unitary}) deserves further attention. With the replacement $\sigma_\pm \rightarrow \alpha^{(*)}$, the term proportional to $\Lambda$ generates a displacement of the resonator field. Moreover, the term proportional to $\xi$ generates squeezing of the field, with a qubit-state dependent squeezing parameter $\xi$. We can thus expect the qubit-resonator state to display the properties of displaced-squeezed states, both transformations being qubit-state dependent. With $\xi=g\Lambda/2\omega_r$, squeezing is expected to be larger for  $\omega_r\ll\omega_a$~\cite{PhysRevA.81.042311}.

\section{Master equations \label{sec-master-equations}}

In this section, we introduce dissipation following two approaches. First, we follow the standard approach where the qubit-resonator coupling $g$ is ignored when obtaining the master equation~\cite{haroche:2006a}. This results in the standard quantum optics master equation~\cite{walls-milburn}. We then consider an approach taking into account the non-negligible qubit-resonator coupling. In both cases, the qubit and the resonator are assumed to be weakly coupled to a bath of harmonic oscillators, with free Hamiltonian 
\be	\label{eqn-baths}
	H_B=\sum_l \nu_lb_l^\dagger b_l,
\ee
where $b_l,b_l^\dagger$ are ladder operators for bath mode $l$ with frequency $\nu_l$ and system-bath coupling
\be	\label{eqn-relaxation-coupling}
	H_\mrm{SB}=\sum_l \alpha_l(c+c^\dagger)(b_l+b_l^\dagger),
\ee
with $\alpha_l$ a coupling strength to bath mode $l$.  For the qubit $c\rightarrow\sigma_-$, while for the resonator $c\rightarrow a$.  In the standard approach, this will correspond to qubit and resonator damping, respectively. Finally, dephasing is modeled classically as
\be	\label{eqn-dephasing-coupling}
	H_\mrm{dep}= f(t) \sz,
\ee
where $f(t)$ is a random function of $t$ with zero mean value. A quantum model for dephasing leads to similar results and is presented in Appendix~\ref{sec-quantum-dephasing} for completeness. As will be seen in Section~\ref{sec-ultrastrong}, though the master equations obtained in the quantum and the classical cases have the same form, asymmetric noise spectral densities are allowed in the quantum model, yielding different predictions in the ultrastrong coupling regime.

\subsection{Standard master equation \label{sec-std}}

The standard approach is to assume that qubit and resonator are independent when obtaining the dissipative part of the master equation. The coupling is then reintroduced in an ad-hoc fashion in the Hamiltonian. This leads to the standard master equation
\be	\label{eqn-std}
	\drv{\rho}{t}=-i\commut{H}{\rho}+\mathcal{L}_\mrm{std}\rho,
\ee
where, at $T=0$,
\be	\label{eqn-std-lindbladian}
	\mathcal{L}_\mrm{std}\cdot = \kappa \mathcal{D}[a]\cdot+\gamma_1\mathcal{D}[\sm]\cdot+\frac{\gamma_\phi}{2}\mathcal{D}[\sz]\cdot,
\ee
with $\mathcal{D}[O]\rho=\frac12\left(2O\rho O^\dagger-\rho O^\dagger O - O^\dagger O \rho\right)$. Here, $\kappa$ is the photon leakage rate for the resonator, $\gamma_1$ the qubit relaxation rate and $\gamma_\phi$ the qubit pure dephasing rate. This expression is obtained in the Markov approximation which assumes the spectral density of all three baths to be white. In other words, the environment-system coupling is evaluated at the relevant frequency ($\omega_r$ for $\kappa$, $\omega_a$ for $\gamma_1$ and $\omega\rightarrow0$ for $\gamma_\phi$) and then assumed to have support at all frequencies. 

For $g/\Sigma$ small enough for the RWA to be safely performed, this expression while not rigorous~\cite{scala2007cavity,PhysRevA.65.023807} can be used to accurately describe many cavity QED and circuit QED experiments~\cite{haroche:2006a,wallraff2004strong}. Indeed, the terms $\mathcal{D}[a]\cdot$ and $\mathcal{D}[\sm]\cdot$ in Eq.~\eq{eqn-std-lindbladian} correctly tend to bring the system to the ground state $\ket{g0}$ of the Jaynes-Cummings Hamiltonian. 

\begin{figure}
	\begin{center}
		\includegraphics[scale=0.7]{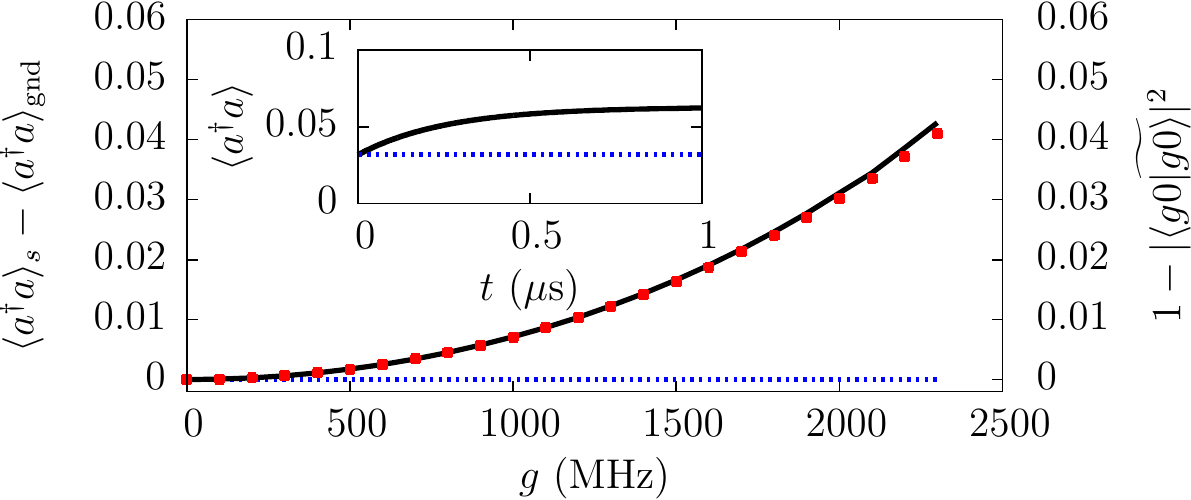}
		\caption{(Color online) Excess in the mean photon number due to relaxation in the steady state of the ultrastrong qubit-resonator system. Initially, the system is in its true ground state $\widetilde{\ket{g0}}$, but, under the standard master equation \eq{eqn-std}, relaxation unphysically excites the system even at $T=0$. The black line, which corresponds to the left axis, represents the number of additional photons introduced in steady state by dissipation. The red dots, associated to the right axis, designate one minus the fidelity of the Rabi ground state $\widetilde{\ket{g0}}$ to the vacuum state $\ket{g0}$. The parameters are $\omega_a/2\pi=\omega_r/2\pi=6$ GHz, $\kappa/2\pi=\gamma_1/2\pi=0.1$ MHz and no pure dephasing. Inset: mean photon number as a function of time for the system starting in its ground state with $g/2\pi=2$ GHz. In both the main plot and the inset, the blue dashed line indicates results for the fidelity and the photon number as obtained with the master equation presented in Sec.~\ref{sec-model}}
		\label{fig-relax}
	\end{center}
\end{figure}

In the ultra-strong coupling regime however, $\ket{g0}$ is no longer the ground state and Eq.~\eq{eqn-std} will bring the ultrastrongly coupled qubit-resonator system outside of its true ground state $\ket{\widetilde{g0}}$. Therefore, even at $T=0$, in which case no energy should be added to the system, relaxation will generate photons in excess to those already present in the ground state. These additional excitations are plotted in Fig.~\ref{fig-relax} as a function of $g$ as the black line. This curve closely follows the behavior of the error one makes by approximating the Rabi ground state by the vacuum state, which is represented by the red dots. It is important to emphasize that these results are obtained for an undriven system evolving simply under dissipative dynamics. We also note that, in practice, preparing $\ket{g0}$ can be extremely challenging, requiring for example tuning of the coupling constant $g$ in a time scale $\ll 1/g$, this being typically in the sub-nanosecond range. This is why  the system is initialized in $\ket{\widetilde{g0}}$ in Fig.~\ref{fig-relax}.

Moreover, because it assumes white noise, Eq.~\eq{eqn-std-lindbladian} incorrectly describes Purcell decay, which can be the factor limiting coherence in superconducting qubits~\cite{PhysRevLett.101.080502}. Additionally, Purcell decay is probing the resonator bath at the qubit transition frequency, something which is missing from the above description but is central to the experiments reported in Refs.~\cite{reed2010fast,houck2007generating} For the same reason and as discussed below, it also incorrectly describes dressed-dephasing~\cite{PhysRevA.77.060305,PhysRevA.79.013819}.

\subsection{Master equation in the dressed picture \label{sec-model}}	

We now take into account qubit-resonator coupling when deriving the master equation. In this case, we cannot assign a unique dissipation channel to each bath mentioned above. Indeed, rather than transitions between eigenstates of the free Hamiltonian $H_0 = \omega_r a^\dag a + \omega_a \sigma_z/2$, coupling to the baths leads to transitions between the qubit-resonator entangled eigenstates $\{\ket{\widetilde{n,\pm}},\ket{\widetilde{g0}}\}$. To simplify the notation, these states will be denoted below as $\ket{j}$, $j$ increasing with energy. These states can be approximated analytically as explained in Sec.~\ref{sec-hamiltonians} or found exactly numerically~\cite{PhysRevB.72.195410, PhysRevLett.99.173601, PhysRevA.80.033846, PhysRevA.82.062320, PhysRevA.82.025802}.

To obtain a master equation that takes into account the coupling $g$, we first move to the frame that diagonalizes the Rabi Hamiltonian for both the system and the system-bath Hamiltonians. Neglecting high-frequency terms, the resulting expressions involve transitions $\ket{j}\leftrightarrow\ket{k}$ between eigenstates at a rate which depends on the noise spectral density at frequency $\Delta_{kj}=\omega_k-\omega_j$. If their linewidth is small enough, these transitions can be treated as due to independent baths.  As a result, these independant baths can each be treated in the Markov approximation~\cite{carmichael}. As shown in Appendices~\ref{sec-master-relax} and~\ref{sec-master-deph}, this leads at $T=0$, to the Lindbladian
\begin{align} \label{eqn-lindbladian}
		\mathcal L_\mrm{dr}\cdot&=\mathcal{D}\left[\sum_j\Phi^{j}\proj{j}\right]\cdot+\sum_{j, k\neq j}\Gamma_\phi^{jk}\,\mathcal{D}\big[\ket{j}\bra{k}\big]\cdot\notag\\
			&\;\;\;+\sum_{j, k>j}\left(\Gamma_\kappa^{jk}+\Gamma_\gamma^{jk}\right)\,\mathcal{D}\big[\ket{j}\bra{k}\big]\cdot,
\end{align}
where $\ket{j}$ and $\ket{k}$ are eigenstates of the qubit-resonator system. Temperature dependence is taken into account in the Appendices but dropped here to simplify the discussion. The first two terms in Eq.~\eq{eqn-lindbladian} are the contributions from the bath described by Eq.~\eq{eqn-dephasing-coupling} that caused only dephasing in the standard master equation. Here, this $\sz$ bath causes dephasing in the eigenstate basis with
\be	\label{eqn:phi_j}
	\Phi_j=\sqrt{\frac{\gamma_\phi(0)}{2}}\sz^{jj},
\ee
where $\gamma_\phi(\omega)$ is the rate corresponding to the dephasing noise spectral density at frequency $\omega$ and where
\begin{equation}\label{eqn-sz}
	\sz^{jk}=\bra{j}\sz\ket{k}.
\end{equation}
Since $\sigma_z$ is not diagonal in the eigenbasis, it also causes unwanted transitions at a rate
\be\label{eq:GammaDressedDephasing}
	\Gamma_\phi^{jk}=\frac{\gamma_\phi(\Delta_{kj})}{2}\times\left|\sz^{jk}\right|^2.
\ee
This contribution will only be significant if the dephasing bath has spectral weight at the potentially high frequency $\Delta_{kj}$ or if the qubit is operated away from a sweet-spot, in which case even low spectral weight can have a large impact~\cite{PhysRevLett.105.100504}. Finally, the last two terms of Eq.~\eq{eqn-lindbladian} are  contributions from the resonator and qubit baths that caused energy relaxation in the quantum optical master equation. They now cause transitions between eigenstates at rates
\begin{align}
	\Gamma_\kappa^{jk}&=\kappa(\Delta_{kj})\times\left|X^{jk}\right|^2\label{eqn:Gammakappa}\\
	\Gamma_\gamma^{jk}&=\gamma(\Delta_{kj})\times\left|\sx^{jk}\right|^2,\label{eqn:Gammagamma}
\end{align}
where
\begin{align}
	X^{jk}&=\bra{j}X\ket{k}\label{eqn-X}\\
	\sx^{jk}&=\bra{j}\sx\ket{k}\label{eqn-sx}.
\end{align}
Here, $\kappa(\omega)$ and $\gamma(\omega)$ are rates that are proportional to noise spectra, respectively for resonator and qubit environments.

The dressed Lindbladian $\mathcal L_\mrm{dr}\cdot$  solves the problem stated in Section~\ref{sec-std}. Indeed, at $T=0$, rather than exciting the system, dissipators accounting for relaxation in Eq.~\eq{eqn-lindbladian} lead to decay to the true ground state. This is illustrated by the dashed blue line in Fig.~\ref{fig-relax}. Moreover, it is interesting to point out that, in addition to the zero-frequency term  responsible for pure dephasing, the noise along $\sz$ can stimulate transitions between the eigenstates $\ket j$, leading to dephasing-induced generation of photons and qubit excitations~\cite{dodonov2009photon}. This is related to the time-dependent Casimir effect, as will be discussed further in Section~\ref{sec-casimir}.

\begin{figure}
	\begin{center}
		\includegraphics[scale=0.5]{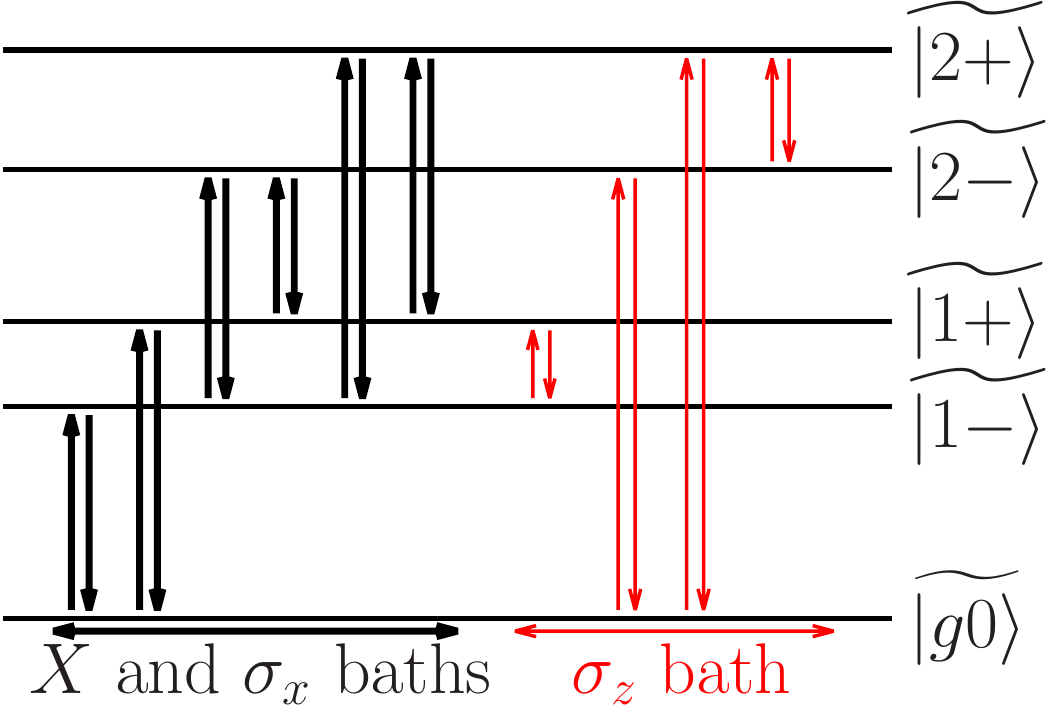}
		\caption{(Color online) Transitions driven by noise. $X$ and $\sigma_x$ baths can only generate transitions between states of different parity. The $\sigma_z$ bath can generate transitions between any pair of levels of same parity.
\label{fig-transitions}}
	\end{center}
\end{figure}
Finally, Fig.~\ref{fig-transitions} illustrates the allowed transitions given the symmetry of the Rabi Hamiltonian, and in particular given that it preserves the parity of the total number of excitation. As further explained in Appendix~\ref{sec-master-relax}, for odd (parity-changing) transition matrices, such as relaxation-related operators $X$ and $\sx$, no decay is possible between states of same parity. On the other hand, the even (parity-preserving) $\sz$ matrix associated with dephasing can generate transitions only between pairs of states of same parity.

\section{Strong coupling regime \label{sec-strong-coupling}}

Before going to the ultra-strong coupling regime, in this section we focus on the simpler strong-coupling regime described by the Jaynes-Cummings Hamiltonian.  We first consider the dispersive regime, a situation which is particularly useful for qubit readout in circuit QED, and then move to the Jaynes-Cummings Hamiltonian. The analysis done here is in the spirit of the dressed-dephasing model~\cite{PhysRevA.79.013819}, but also encompasses qubit-resonator resonance~\cite{PhysRevA.75.013811}.

\subsection{The dispersive regime \label{subsec:dispersive}}

As already mentioned above, due to the white noise approximation, the description provided by the standard master equation of Sec.~\ref{sec-std} can break down even when the dispersive approximation is valid. In this regime, achieved when $|\Delta| =|\omega_a-\omega_r| \gg  g$, the Jaynes-Cummings Hamiltonian reduces to the dispersive Hamilonian
\be	\label{eqn-disp}
	H_\mrm{disp}=(\omega_r+\chi\sz+\zeta)\ad a+\frac{\omega_a+\chi}{2}\sz+\zeta\sz(\ad a)^2,
\ee
to fourth order in $g$. This diagonal Hamiltonian includes an effective qubit-resonator dispersive coupling $\chi=g^2/\Delta$ and a small nonlinearity $\zeta=g^4/\Delta^3$ which is usually neglected. The coupling implies that the dispersive eigenstates display some degree of mixing between qubit and resonator. Indeed, to second order in $g$, these eigenstates, denoted $\overline{\ket{gn}}$ and $\overline{\ket{en}}$, are
\begin{align}\label{eq:DisperssiveDressedStates}
	\overline{\ket{e,n-1}}&\simeq\left(1-\frac{g^2n}{2\Delta^2}\right)\ket{e,n-1}-\frac{g\sqrt{n}}{\Delta}\ket{g,n}\\
	\overline{\ket{g,n}}&\simeq\left(1-\frac{g^2n}{2\Delta^2}\right)\ket{g,n}-\frac{g\sqrt n}{\Delta}\ket{e,n-1}.
\end{align}

Consequences of this mixing of qubit and resonator states are Purcell decay~\cite{PhysRev.69.37} and the dressed-dephasing model~\cite{PhysRevA.77.060305, PhysRevA.79.013819}. Purcell decay is the relaxation of the qubit by photon emission out of the cavity. This is captured here by the rate $\Gamma_\kappa^{jk}$ evaluated between the dressed states $\overline{\ket{e,0}}$ and $\overline{\ket{g0}}$ yielding $\Gamma_\kappa^{\overline{e0},\overline{g0}} = (g/\Delta)^2 \kappa(\omega_a+\chi)$.  In this expression, the standard Lindbladian Eq.~\eq{eqn-std-lindbladian} rather evaluates cavity damping at the cavity frequency $\omega_r$. This difference is important in several circuit QED experiments~\cite{reed2010fast,houck2007generating,PhysRevLett.101.080502}.

The standard approach also does not capture correctly dressed-dephasing discussed in Refs.~\cite{PhysRevA.77.060305, PhysRevA.79.013819}. Essentially, dressed-dephasing captures how dephasing can produce relaxation because of the finite qubit-photon mixing in the dressed states Eq.~\eq{eq:DisperssiveDressedStates}. Deriving the dressed-dephasing rate from Eq.~\eq{eqn-std-lindbladian} yields a result proportional to the spectrum of dephasing noise at zero frequency. Here and in Refs.~\cite{PhysRevA.77.060305, PhysRevA.79.013819}, we rather obtain the rate $\Gamma_\phi^{jk}$ involving the spectrum of dephasing noise at the qubit-resonator detuning frequency. Assuming a dephasing noise scaling as $1/f$, the difference between the two predictions can be quite large. In practice, this means that one must be careful in interpreting results of numerical simulations of the standard Lindbladian~Eq.~\eq{eqn-std} as it can include unrealistically large amounts of qubit flipping induced by dephasing noise. It is worth pointing out however that dressed-dephasing can be relevant experimentally in some circumstances~\cite{PhysRevLett.105.100504}.

In Refs.~\cite{PhysRevA.77.060305, PhysRevA.79.013819}, the nonlinear term proportional to $\zeta$ is neglected in the derivation of the master equation containing the dressed-dephasing contribution. This approximation breaks down when $\zeta>\kappa$ in which case the approach developed here is appropriate. However, when $\zeta \sim \kappa$, the different resonator transitions are not well separated and the environment cannot be treated as independent baths. The approach of Refs.~\cite{PhysRevA.77.060305, PhysRevA.79.013819} should then be used. The validity of the results obtained here is further discussed in Appendix~\ref{sec:conditions}.

\subsection{Dissipation in the Jaynes-Cummings model \label{subsec:JC}}

In this section, we consider the situation where the dispersive approximation does not hold but the RWA is still valid. This is essentially generalizing the results of the dressed-dephasing model~\cite{PhysRevA.77.060305, PhysRevA.79.013819}. Under the RWA, the ground state is simply $\ket{g0}$. Excited eigenstates $\ket{n\pm}$ are given by Eqs.~(\ref{eqn-nplus}) and~(\ref{eqn-nminus}), with the mixing angle $\theta_n$ defined by Eq.~(\ref{eqn-angle}) and $\mu=0$ such that $\Delta_n^\mrm{BS}=\omega_a-\omega_r = \Delta$. 

\begin{figure}
	\begin{center}
		\includegraphics[scale=0.8]{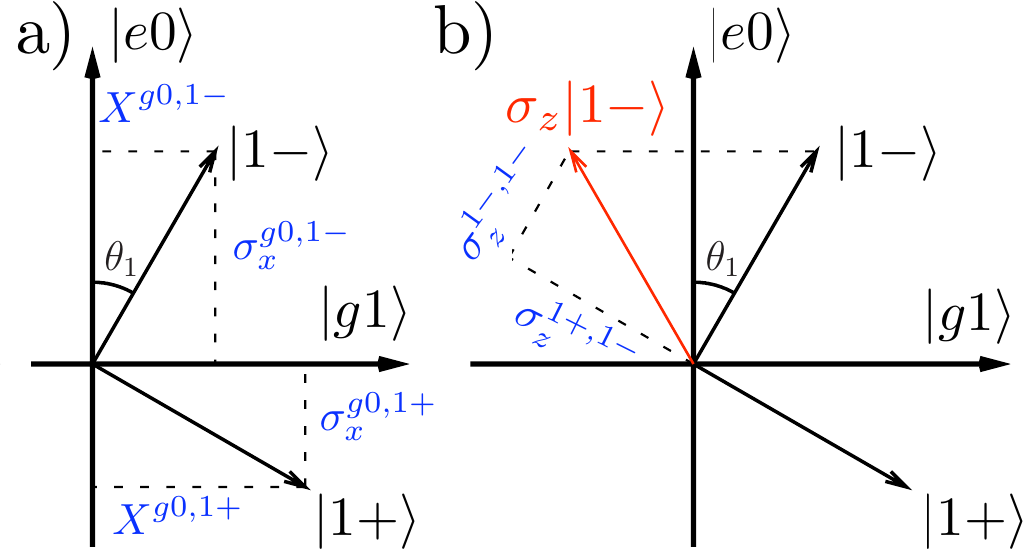}
		\caption{(Color online) Matrix elements under the RWA in  the subspace $\{\ket{g0},\ket{1-},\ket{1+}\}$.  a) Relaxation matrix elements. Eigenstates are a mixture of qubit and resonator states, with an angle $\theta_1$. The fraction of the relaxation rate that comes from the qubit or the resonator bath is determined by the projection of the eigenstate on the qubit $\ket{e0}$ or resonator $\ket{g1}$ axis. b) Dephasing matrix elements. The dephasing Hamiltonian rotates state vectors around the $\ket{e0}$ axis. Resulting vectors have a projection on the orthogonal eigenstate in the same doublet. This generates transitions between $\ket{1,+}$ and $\ket{1,-}$ if $\theta_1>0$.\label{fig-mixing}}
	\end{center}
\end{figure}

We first consider the matrices $X$ and $\sx$, whose elements are involved in relaxation rates described by Eqs.~\eq{eqn:Gammakappa} and~\eq{eqn:Gammagamma}. To keep the discussion simple, we limit ourselves to the subspace $\{\ket{g0},\ket{1-},\ket{1+}\}$. Complete results can be found in Appendix~\ref{sec-elements}. Since the Jaynes-Cummings eigenstates have a well-defined excitation number, only transitions involving the loss or gain of  one quantum are allowed, thus forbidding transitions between $\ket{1+}$ and $\ket{1-}$. This yields
\begin{align}
	X = &\mattrois{0&\sin\theta_1&\cos\theta_1\\\sin\theta_1&0&0\\\cos\theta_1&0&0}\\
	\sx = &\mattrois{0&\cos\theta_1&-\sin\theta_1\\\cos\theta_1&0&0\\-\sin\theta_1&0&0}.
\end{align}
With the eigenstates changing character with $\theta_1$ between mostly qubit-like or photon-like, the contribution of the two decay channels $X$ and $\sx$ follows. This can be visualized geometrically as illustrated in Fig.~\ref{fig-mixing}a. 
In particular, when the qubit and the resonator are on resonance, their corresponding relaxation noises have exactly the same weight. For example, the matrix elements of $X$ reduce to
\begin{align}
	X^{g0;1,\pm}&=\pm\frac{1}{\sqrt{2}}\\
	X^{n,+;n+1,+}&=X^{n,-;n+1,-}=\frac12\left(\sqrt{n}+\sqrt{n+1}\right)\\
	X^{n,+;n+1,-}&=X^{n,-;n+1,+}=\frac12\left(\sqrt{n}-\sqrt{n+1}\right),
\end{align}
which exactly leads to the master equation presented in Ref.~\cite{PhysRevA.75.013811} in the presence of resonator losses only.

Under the RWA, $\sz^{jk}$ can only be non-zero for states that involve the same total excitation number, i.e. are in the same Jaynes-Cummings doublet. The resulting matrix elements are
\begin{align}
	\sz^{g0;g0}&=-1\\
	\sz^{n\pm;n\pm}&=\mp\cos(2\theta_{n})\\
	\sz^{n\mp;n\pm}&=-2\cos\theta_n\sin\theta_n.
\end{align}
Generalizing the dressed-dephasing model, the above formulae show that the dephasing bath induces transitions between states in the same JC doublet. As illustrated in Fig.~\ref{fig-mixing}b, this happens only if there is some mixing between the qubit and the resonator. In particular, in resonance, $\sz^{n\pm;n\pm}=0$ and $\sz^{n\mp;n\pm}=1$. Then, dephasing processes for states that do not involve $\ket{g0}$ are entirely due to transitions within the doublets which are caused by dephasing noise at the doublet splitting frequencies $2g\sqrt{n}$. Since these are very high frequencies and dephasing is often caused by a $1/f$ bath~\cite{PhysRevLett.97.167001}, transition rates within doublets are expected to be small. Therefore, in resonance, states that do not contain $\ket{g0}$ should be largely immune to pure phase-destroying processes.

\section{Ultrastrong coupling regime \label{sec-ultrastrong}} 

Here, we take the ratio $g/\Sigma$ to be sufficient to break the RWA, but still much smaller than unity. In this situation, the transition matrix elements given in Eqs.~(\ref{eqn-sz}), (\ref{eqn-X}), and (\ref{eqn-sx}) can be evaluated using the Bloch-Siegert eigenstates Eqs.~(\ref{eqn-nplus}) and (\ref{eqn-nminus}), as done in Appendix~\ref{sec-elements}. In this section, we use these results to study two distinctive phenomena occuring in the ultrastrong coupling regime: 1) asymmetry of the vacuum Rabi splitting spectrum; 2) sideband transitions and photon generation caused by qubit frequency modulations.

\subsection{Asymmetry of the vacuum Rabi splitting spectrum \label{sec-asymmetry}}

Vacuum Rabi splitting is observed by measuring transmission ($\mathbb{I}\mrm{m}\mean{a}$ and/or $\mathbb{R}\mrm{e}\mean{a}$) of the resonator under weak rf excitation~\cite{wallraff2004strong, fink2008climbing}. In the presence of the rf drive, the Hamiltonian becomes
\be	\label{eqn:drive}
	H_\mrm{drvn}(t)= H_R+\epsilon\,a\eul{i\nu t} + \epsilon^\ast\,a^\dagger\eul{-i\omega_d t},
\ee
with $\epsilon$ the amplitude of the drive and $\omega_d$ its frequency. Assuming  $g\ll\Sigma$, we find in Appendix~\ref{sec-rabi} under the three-level approximation that $\mathbb{I}\mrm{m}\mean{a}$ in steady-state and at the Bloch-Siegert-shifted qubit-resonator resonance ($\Delta^{\mrm{BS}} = 0$) is given by 
\be\label{eq:IMaBS}
	\mathbb{I}\mrm{m}\mean{a}_s=-\frac{\epsilon\,\Gamma_1/2}{\Gamma_1^2+(\Delta^\mrm{BS}+g)^2}-\frac{\epsilon\,\Gamma_2/2}{\Gamma_2^2+(\Delta^\mrm{BS}-g)^2},
\ee
\begin{figure}
	\begin{center}
		\includegraphics[scale=0.65]{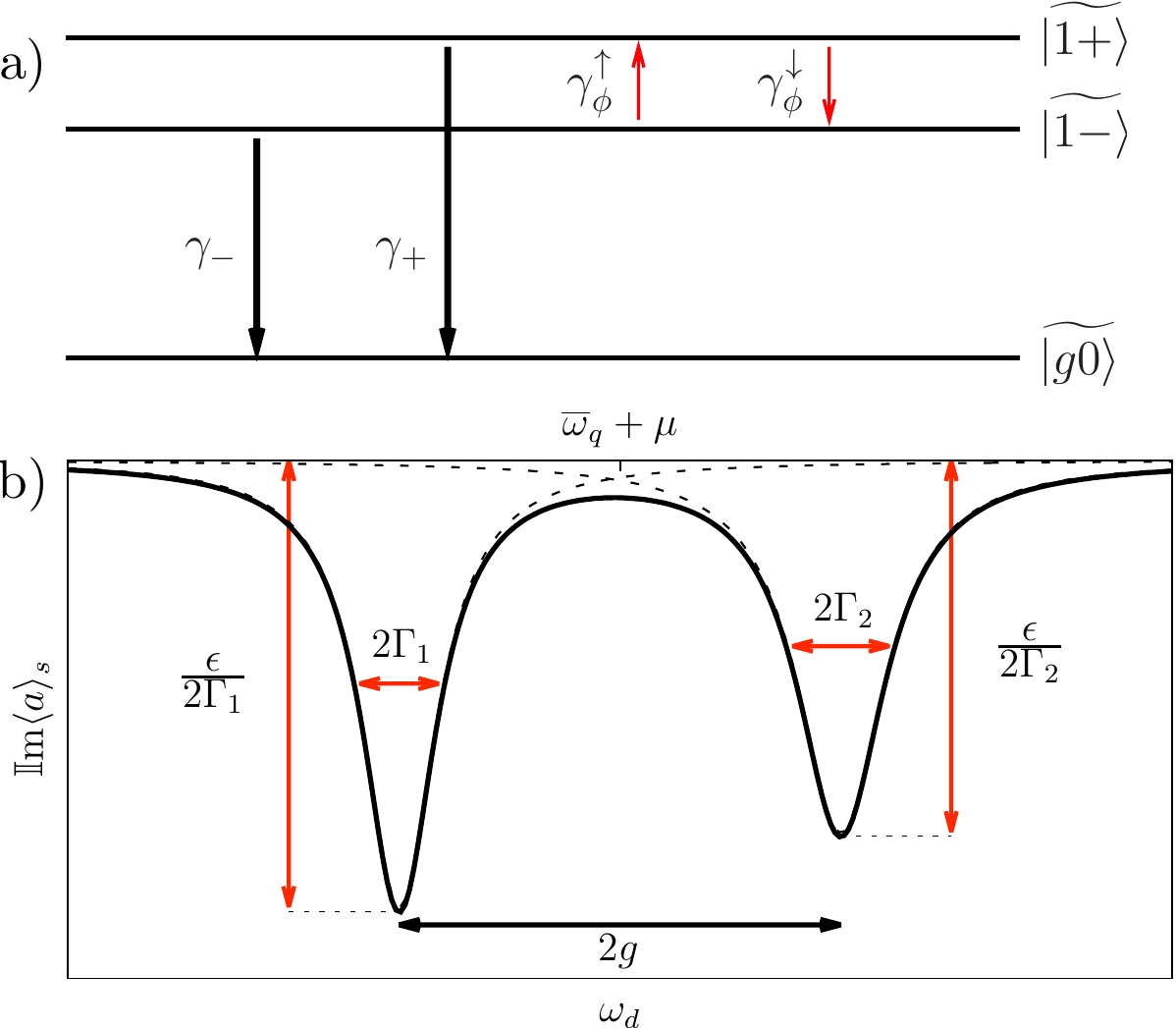}
	\end{center}
	\caption{(Color online) Vacuum Rabi splitting. a) Transition rates involved in the perturbative calculation in the three-level approximation. b) Schematic plot of $\mathbb{I}\mrm{m}\mean{a}_s$ as a function of $\omega_d$. In general, the result is not symmetric.}\label{fig-ima}
\end{figure}
where
\begin{align}
	\Gamma_1&=\frac12\left(\gamma_-+\gamma_\phi^\uparrow+\gamma_\phi^-\right)\\
	\Gamma_2&=\frac12\left(\gamma_++\gamma_\phi^\downarrow+\gamma_\phi^+\right).
\end{align}
The various rates entering these expressions are illustrated in Fig.~\ref{fig-ima}a) and can be found in Appendix~\ref{sec-rabi}. As in the standard case, the transmission is composed of two Lorentzians separated by $2g$~\cite{haroche:2006a}. However, here two distinct rates $\Gamma_1$ and $\Gamma_2$ dictate the width and height of these peaks. As a result, the vacuum Rabi splitting spectrum can be asymmetric, even when the qubit and the resonator are in resonance. Asymmetry in the presence of counter-rotating terms has also been pointed out considering only cavity decay in the Fourier transfom of $\langle\sz(t)\rangle$~\cite{hausinger2008dissipative} and qubit fluorescence~\cite{2010arXiv1009.4366C}.

Here, three situations can lead to asymmetry:
\begin{enumerate}
	\item Relaxation noise spectra are not equal at the frequencies corresponding to the two transitions $\ket{\widetilde{1\pm}}\rightarrow\ket{\widetilde{g0}}$. This situation will be referred to as the \emph{non-white relaxation noise} case.
	\item The pure qubit dephasing noise spectrum is not equal at frequencies $\Delta_{1\pm,1\mp}$. Since classical noise spectra are always symmetric in frequency~\cite{clerk2010introduction}, we call this situation the \emph{quantum dephasing noise} case.
	\item Keeping counter-rotating terms such that $\Lambda\neq0$, the matrix elements of $X$ and $\sigma_x$ for transitions $\ket{\widetilde{1\pm}}\rightarrow\ket{\widetilde{g0}}$ and $\ket{\widetilde{1\pm}}\rightarrow\ket{\widetilde{1\mp}}$ are not equal, as shown in appendix~\ref{sec-elements}. This is the \emph{ultrastrong} case.
\end{enumerate}

To characterize the asymmetry as the coupling $g$ is increased, the noise spectra must be known. We now make some hypotheses on that noise and consider their consequences. We first isolate the influence of counter-rotating terms by choosing white relaxation noise and no pure dephasing. In these conditions, the second order terms in $g$ cancel out and the asymmetry $\eta = \Gamma_1-\Gamma_2$ increases linearly with $g$:
\be	\label{eqn:etaUS}
	\eta_\mathrm{us} = \frac{\Lambda}{2}(\kappa+\gamma_1).
\ee
For the parameters realized in Ref.~\cite{niemczyk2010circuit}, $g/2\pi=636$~MHz, $\omega_r/2\pi=5.357$~GHz, and $\kappa/2\pi=3.7$~MHz, and taking $\gamma_1/2\pi=0.1$ MHz yields $\eta/2\pi \sim 0.11$ MHz and in turn an asymmetry of $\sim 6\%$ in the transmission peak amplitudes. As a result, in the ultrastrong regime, the height of the transmission peaks in a vacuum Rabi splitting experiment cannot be used to tune the qubit and the resonator exactly in resonance.

In general however, noise is not white. Though the ohmic model which leads to constant relaxation rates $\kappa(\omega)$ and $\gamma(\omega)$ is usually valid, the transition rates $\gamma_\phi^{\uparrow/\downarrow}$ coming from the dephasing bath can be asymmetric. This yields a contribution $\eta_\phi$ to the total asymmetry $\eta=\eta_\mathrm{us}+\eta_\mathrm{\phi}$, where $\eta_\mathrm{us}$Ó is given by equation~\eq{eqn:etaUS} and
\be
	\eta_\phi=\frac{1-4\Lambda^2}{8}\left[\gamma_\phi(\Delta_{1-,1+})-\gamma_\phi(\Delta_{1+,1-})\right].
\ee
As discussed in Appendix~\ref{sec-quantum-dephasing}, noise at negative frequencies will only appear for non-zero effective bath temperatures, i.e.~the rates respect detailed balance. With $\Delta_{1\pm,1\mp}=\pm2g$, we therefore obtain
\be
	\gamma_\phi(-2g)=\exp\left(-\frac{2g}{k_B T}\right)\gamma_\phi(2g),
\ee
and have 
\be
	\eta_\phi \simeq \frac{1-4\Lambda^2}{8}\left[\exp\left(-\frac{2g}{k_B T}\right)-1\right]\gamma_\phi(2g).
\ee
We now distinguish two  limits. If $k_B T\gg2g$, $\eta_\phi\rightarrow0$ and we retrieve the classical noise limit. The asymmetry is then entirely due to the ultrastrong coupling. It is possible to isolate this ultrastrong signature by increasing the effective temperature of the bath, for example in circuit QED, by injecting noise in $\omega_a$ with an external flux line. In the opposite scenario, if $ k_BT \ll 2g$, the asymmetry $\eta_\phi$ becomes important. In particular, if $T\rightarrow0$ 
\be
	\eta_\phi \simeq \frac{1-4\Lambda^2}{8}\gamma_\phi(2g).
\ee
Knowing the ultrastrong contribution to asymmetry, either by calculating it with equation (39) or by measuring it experimentally, it is possible to isolate the effect of quantum dephasing noise by taking $\eta_\phi=\eta-\eta_\mathrm{us}$. This asymmetry is thus a probe for dephasing noise at the vacuum Rabi splitting frequency.  The ultrastrong coupling regime widens the range of accessible values of $g$. As a result, the noise spectrum entering the rate $\gamma_\phi(\omega)$ could realistically be investigated to frequencies up to $\sim2$ GHz, where data is lacking~\cite{bylander2011noise} and where a crossover from $1/f$ to ohmic behaviour is expected to happen~\cite{PhysRevLett.94.127002}.

\subsection{Qubit frequency modulations: sidebands and photon generation \label{sec-casimir}}	

In this section, we focus on the effect of the term $f(t) \sigma_z$ of the classical dephasing model Eq.~\eq{eqn-dephasing-coupling}. We first consider the case where $f(t)$ is a controlled modulation of the qubit frequency (for example, using an external flux) before turning to the situation where $f(t)$ represents incoherent noise. Both cases will be related to the dynamical Casimir effect~\cite{PhysRevLett.103.147003, PhysRevLett.105.233907}.

For $g\ll\Sigma, \, \Delta$, we apply on $H=H_R+f(t)\sz$ the dispersive transformation~\cite{harocheBookFundamental}, here generalized to take into account the counter-rotating terms~\cite{PhysRevA.80.033846}
\be
	U_\mrm{D}=\exp\left\{\lambda\left(\ad\sm-a\spl\right)+\Lambda\left(a\sm -\spl\ad\right)\right\}.
\ee
To second order in $g$, we find
\begin{align}
	H_D(t) &\simeq H_0' +\chi'(t)\ad a\sz+f(t)\sz\notag
		-2f(t)(\lambda I_++\Lambda I_\mrm{CR})\notag\\
		&-2f(t)\lambda\Lambda\sz(a^2+\ad\,\!^2),
\end{align}
where $\chi'(t)=-2(\lambda^2+\Lambda^2)f(t)$ and $H_0' = [\omega_r+(\chi+\mu)\sz]a^\dag a+ [\omega_a + \chi+ \mu] \sigma_z/2$ the free but Lamb and Bloch-Siegert-shifted Hamiltonian.

 We first focus on the case of a classical modulation $f(t)=\epsilon_z \cos\omega_dt$ of the qubit transition frequency. For $\epsilon_z\ll\omega_d$, the oscillating terms proportional to $\sz$ and $\ad a\sz$ can be dropped under the RWA. Depending on the choice of modulation frequency $\omega_d$, it is possible to select different terms in $H_D(t)$ while dropping others. First, for $\omega_d=\Delta$, we have $H_D(t) \simeq H_0' -\epsilon_z\lambda I_+$ corresponding to a red sideband transition.  For $\omega_d=\Sigma$, we rather find $H_D(t) \simeq H_0' -\Lambda\epsilon_zI_\mrm{CR}$ corresponding to a blue sideband transition. Interestingly, these sideband rates are in first rather than second order in $g/\omega_d$. This is to be contrasted to the usual second order results obtained in circuit QED~\cite{blais2007quantum,wallraff2007sideband,PhysRevB.79.180511} and could be used to speed up two-qubit gates. Finally, modulating at twice the resonator frequency,  $\omega_d = 2\omega_r$, the Hamiltonian reduces to $H_D(t)= H_0' -\epsilon_z\lambda\Lambda\sz(\ad\,\!^2+a^2)$, corresponding to a pumped parametric oscillator~\cite{walls-milburn}. Rather than modulating the resonator frequency~\cite{yamamoto:2008a}, here it is the qubit that acts as a moving boundary condition. In Ref.~\cite{dodonov2009photon} this Hamiltonian was also linked to the  dynamical Casimir effect.

\begin{figure}
	\begin{center}
		\includegraphics[scale=0.7]{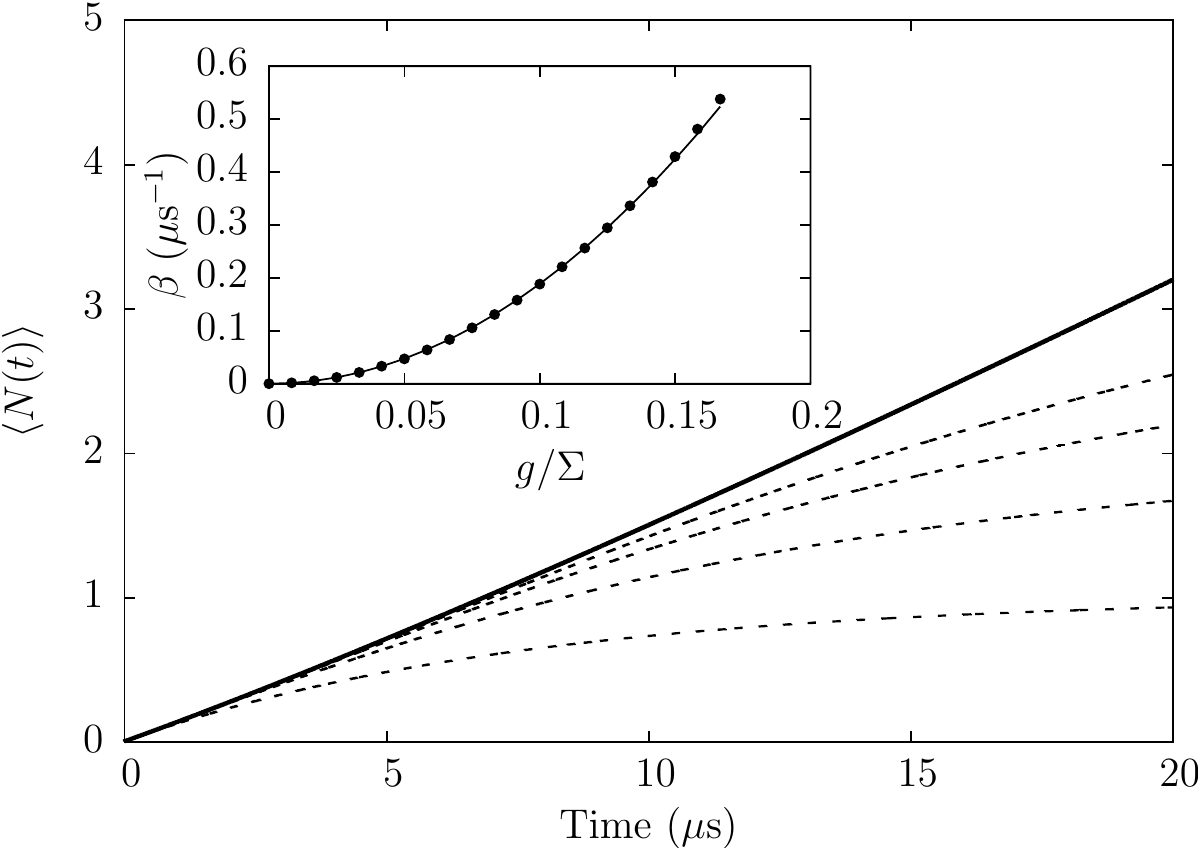}
		\caption{Photon generation due to dephasing with the Lindbladian Eq.~\eq{eqn-lindbladian}. Full line~: white noise. Dotted line~: white noise with a cut-off frequency increasing from bottom to top. For the bottom dotted line the cut-off is such that only transitions up to $\widetilde{\ket{i,\pm}}$ for $i=2$ are driven. For the top curve, transitions from $\widetilde{\ket{g0}}$ through $i=8$ are driven. Inset~: photon generation rate $\beta$ as a function of $g$ for white noise. Points~: numerical results. Line~: perturbation theory Eq.~\eq{eqn-rate}. The parameters are $\omega_a/2\pi=\omega_r/2\pi=6$ GHz, $g/2\pi=1$ GHz and $\gamma_\phi/2\pi=1$ MHz.}
		\label{fig-decoherence}
	\end{center}
\end{figure}

We now move to the situation where $f(t)$ is a random function representing a classical dephasing bath whose spectral content may contain one or more of the above-mentioned relevant frequencies. If the spectral content extends to very high frequencies, it may act on the system through a combination of the above blue and red sideband transitions and photon-pair production, bringing it to an excited state which may display some degree of squeezing. While this discussion is only valid in the dispersive regime, we can extend these results to arbitrary ratios $g/\Delta$. Indeed, in general the $\sigma_z$ bath can drive any transition between Rabi eigenstates of same parity, as illustrated in Fig.~\ref{fig-transitions}. If the corresponding frequencies are present in $f(t)$, combinations of qubit and resonator excitations are produced. In the simplest case where relaxation is neglected and the dephasing bath is white, this leads to a photon creation rate $\beta$. As shown in appendix~\ref{sec-rate},  to second order in~$g$,
\be	\label{eqn-rate}
	\beta = 2 \gamma_\phi\Lambda^2\;T(\theta_2),
\ee
where we have defined $T(\theta_2)=1+2\cos^2\theta_2\sin^2\theta_2$.
This expression is compared to exact numerics in the inset of Fig.~\ref{fig-decoherence}.  It analytically explains the $\Lambda^2$ dependence of the photon creation rate observed numerically by Werlang et al.~\cite{PhysRevA.78.053805} for the special case of $\omega_a = \omega_r$. In this work, the authors have used the standard Lindladian Eq.~\eq{eqn-std-lindbladian}, assuming white noise and corresponding to the full line in Fig.~\ref{fig-decoherence}. If the noise causing dephasing has a $1/f$ spectrum, the standard Lindbladian therefore greatly exaggerates this effect. If noise decreases at higher frequencies, photon generation has a smaller rate, but should also saturate, as illustrated by the dotted lines in Fig.~\ref{fig-decoherence}. This is again a clear demonstration of the breakdown of the standard approach to treating dissipation in the presence of the counter-rotating terms.

Finally, since the master equation is exactly the same in the quantum treatment of dephasing shown in Appendix~\ref{sec-quantum-dephasing}, the above results remain valid in that case. However, the quantum approach explicitly incorporates  temperature in a way that respects detailed balance. This implies that, at $T=0$, $\gamma_\phi(\omega)=0$ for $\omega<0$. Since, as shown in Appendix~\ref{sec-rate}, this Casimir-like photon generation needs negative frequencies, a quantum dephasing bath could not generate excitations in the system at $T=0$. In this model, photon production through dephasing is thus intrinsically a thermal effect.

\section{Conclusion}

We have shown the importance of treating the qubit-resonator system as a whole when studying its interaction with the environment. In particular, we have  shown that the description offered by the standard master equation can break down, for example producing spurious qubit flipping or photon generation, even at zero temperature. To cure these unphysical problems, we have included the qubit-resonator coupling in the derivation of the master equation. The rates entering the modified master equation then depend on the spectrum of noise evaluated at the dressed transition frequencies. These rates have been obtained analytically for a qubit-resonator coupling $g$ that is large enough for individual qubit and/or resonator transitions to be resolved, and for the dispersive ($|\omega_a-\omega_r|\gg g$) to the Bloch-Siegert ($\omega_a+\omega_r\gg g$) regime. Even when including the counter-rotating terms in the qubit-resonator coupling, the results obtained here can be used beyond these regimes by relying on simple numerical diagonalization of the Rabi Hamiltonian. Results in the ultrastrong coupling regime ($g\sim\omega_a,\,\omega_r$) have been presented.

In our model, noise that caused pure dephasing in the standard master equation can now cause transitions in the system. In this sense, the master equation developed here can be viewed as an extension of the dressed-dephasing model~\cite{PhysRevA.77.060305,PhysRevA.79.013819}. In the Bloch-Siegert regime, we find that the vacuum Rabi splitting spectrum can be asymmetric. This asymmetry can be used as a probe of the dephasing noise spectral density at currently unexplored frequencies $\sim1$ GHz and above. Additionally, modulations of the qubit transition frequency can be used to generate red and blue sidebands, or as a parametric oscillator inducing squeezing. Finally, while this means that noise in $\sigma_z$ can generate photons~\cite{PhysRevA.78.053805, dodonov2010cold,1742-6596-274-1-012137}, our model reasonably shows that these spurious excitations cannot be generated at zero temperature.

\begin{acknowledgments}
We acknowledge J.~Bourassa, M.~Boissonneault, and C.~M\"uller for useful discussions. F.B. was supported by NSERC and FQRNT and A.B. by NSERC, the Alfred P. Sloan Foundation, and CIFAR.
\end{acknowledgments}

\appendix

\section{Dissipators for $X$ and $\sx$ baths \label{sec-master-relax}}

In this appendix, we derive in the dressed basis the Lindbladian corresponding to coupling to the $X$ and $\sx$ baths. We take an arbitrary qubit-resonator system with the only assumption that the total excitation number has a well-defined parity. As stated in section~\ref{sec-master-equations}, we assume that the system is coupled to two independent baths of quantum harmonic oscillators with an interaction of the form given by Eq.~(\ref{eqn-relaxation-coupling}). Focusing here on only one bath, we find that in the interaction picture with respect to the free system and bath Hamiltonians, the coupling takes the form
\be
	H_\mrm{SB}(t)= \sum_l \alpha_l\eul{iH_\mrm{S}t}(c
		+c^\dagger)\eul{-iH_\mrm{S}t}(b_l \eul{-i\nu_l t}+b_l^\dagger\eul{i\nu_l t}).
\ee
Expressing the system Hamiltonian in the dressed basis
\be
	H_\mrm{S}=\sum_j E_j\proj{j},
\ee
we have
\begin{align}
	H_\mrm{SB}(t)&=\sum_{jkl} \alpha_lC_{jk}\ket{j}\bra{k}\left(b_l\eul{-i\nu_l t}+b_l^\dagger\eul{i\nu_l t}\right)\eul{i\Delta_{jk}t}\notag\\
\end{align}
where $C_{jk}=\bra{j}(c+c^\dagger)\ket{k}$ and $\Delta_{jk}=E_j-E_k$. We now split the sum in three parts
\begin{align}
	H_\mrm{SB}(t)=&\sum_{l,j} \alpha_l C_{jj}\proj{j}\left(b_l\eul{-i\nu_l t}+b_l^\dagger\eul{i\nu_l t}\right)\notag\\
		&+\left\{\sum_{l}\sum_{j,k>j}+\sum_{l}\sum_{j,k<j}\right\} \alpha_lC_{jk}\ket{j}\bra{k} \notag\\
			&\;\;\;\;\;\;\;\times\left(b_l\eul{-i(\nu_l-\Delta_{jk})t} + b_l^\dagger\eul{i(\nu_l+\Delta_{jk})t}\right).
\end{align}
Since $C_{kj}=C_{jk}^\ast$, this becomes
\begin{align}
	H_\mrm{SB}(t)&=\sum_{j} \sum_l \alpha_l C_{jj}\proj{j}\left(b_l\eul{-i\nu_l t}+b_l^\dagger\eul{i\nu_l t}\right)\notag\\
		&+\sum_{j,k>j} \sum_l \alpha_lC_{jk}\ket{j}\bra{k} b_l^\dagger\eul{i(\nu_l+\Delta_{jk})t} 
		+ \mathrm{h.c.}
\end{align}
We now introduce the operator $\Pi=(-1)^{\ad a+\spl\sm}$, whose eigenvalues label the parity of the total excitation number in the qubit-resonator system. Since $[H_R,\Pi]=0$, system eigenstates $\ket{j}$ have a well-defined parity. As $c$ and $c^\dagger$ change the excitation number by one, $c+c^\dagger$ flips the parity when applied on a state. Therefore, $C_{jj}=0$, which simplifies the Hamiltonian to
\be	\label{eqn-Hint}
	H_\mrm{SB}(t)= s(t) B^\dagger(t)+s^\dagger(t)B(t),
\ee
where
\begin{align}
	s(t)&=\sum_{j,k>j}C_{jk}\ket{j}\bra{k}\eul{i\Delta_{jk}t}\\
	B(t)&=\sum_l\alpha_l b_l\eul{-i\nu_lt}.
\end{align}
This formulation makes it easy to write the Born master equation for the system. Indeed, following the standard procedure, we find~\cite{carmichael}
\begin{align} \label{eqn-Born}
	&\dot{\rho}_I(t)\!=\!\int_0^t\!\diff t'\!\left[s(t')\rho_I(t')s(t)\!-\!s(t)s(t')\rho_I(t')\right]\!\mean{B^\dagger(t)B^\dagger(t')}\notag\\
		&+\int_0^t\diff t'\left[s^\dagger(t')\rho_I(t')s^\dagger(t)-s^\dagger(t)s^\dagger(t')\rho_I(t')\right]\mean{B(t)B(t')}\notag\\
		&+\int_0^t\diff t'\left[s^\dagger(t')\rho_I(t')s(t)-s(t)s^\dagger(t')\rho_I(t')\right]\mean{B^\dagger(t)B(t')}\notag\\
		&+\int_0^t\diff t'\left[s(t')\rho_I(t')s^\dagger(t)-s^\dagger(t)s(t')\rho_I(t')\right]\mean{B(t)B^\dagger(t')}\notag\\
		&	+ \mathrm{h.c.}
\end{align}
From this point, we make assumptions that are standard in the Born-Markov treatment of dissipation~\cite{carmichael}, except for the following considerations. In each term of the Born master equation, we find oscillating exponentials of the form $\exp[i(\Delta_{jk}-\Delta_{j'k'})t]$. Since $k>j$ and $k'>j'$, the argument of these exponentials will be zero for $j=j'$ and $k=k'$, or for pairs of different transitions in the system occuring at the same frequency. As discussed in Appendix~\ref{sec:conditions}, in practice we are often interested only in a subset of the energy levels of the system for which all transitions have different frequencies. In that case, we can neglect all fast oscillating terms to obtain the following master equation in the Schr\"odinger picture
\begin{align}
	\dot{\rho}(t)&=-i\commut{H_S'}{\rho(t)}\label{eqn-relaxation}\\
		&+\sum_{j,k>j}\Gamma^{jk}\overline{n}(\Delta_{kj},T)\mathcal{D}[\ket{k}\bra{j}])\rho(t)\notag\\
		&+\sum_{j,k>j}\Gamma^{jk}\left(1+\overline{n}(\Delta_{kj},T)\right)\mathcal{D}[\ket{j}\bra{k}]\rho(t),\notag
\end{align}
with $\Gamma^{jk}=2\pi d(\Delta_{kj})\alpha^2(\Delta_{kj})|C_{jk}|^2$ and where we have introduced the density of states $d(\nu)$ of the bath. We have also defined the Lamb-shifted system Hamiltonian
\begin{align}	
	H_S'(t)=H_S-\sum_{jk}|C_{jk}|^2&\left\{L_{jk}\proj{k}\right.\notag\\
		&\left.+L_{jk}'(\proj{k}-\proj{j})\right\}.
\end{align}
$L'_{jk}$ are Lamb shifts caused by coupling to the environement and are given by
\begin{align}
	L_{jk}&=\frac{P}{2\pi}\int_0^\infty\diff\nu\,\frac{\Gamma(\nu)}{\nu+\Delta_{jk}}	\label{eqn-Ljk}\\
	L'_{jk}&=\frac{P}{2\pi}\int_0^\infty\diff\nu\,\frac{\Gamma(\nu)\overline{n}(\nu,T)}{\nu+\Delta_{jk}},	\label{eqn-Ljkprime}
\end{align}
and $P$ is Cauchy's principal value. The function $\Gamma(\nu)$ is a relaxation rate. In the case of photon loss, $c\rightarrow a$ and we replace $\Gamma(\nu)$ by $\kappa(\nu)$. In the case of qubit relaxation, $c\rightarrow \sigma_-$ and we replace $\Gamma(\nu)$ by $\gamma(\nu)$. In the main body of this paper, we neglect these Lamb shifts.

\section{Dissipators for the $\sz$ bath \label{sec-master-deph}}

\subsection{Classical model \label{sec-classical-dephasing}}

In this section, we derive the dephasing part of Eq.~(\ref{eqn-lindbladian}). For this, we introduce a stochastic function $f(t)$ modulating the qubit frequency
\be\label{eq:ClassicalDephasingNoise}
	H_\mrm{dep}= f(t) \sz,
\ee
where the mean value of $f(t)$ vanishes. Following Appendix~\ref{sec-master-relax}, we express the Hamiltonian in the dressed basis and move to the interaction picture with respect to Eq.~\eq{eq:ClassicalDephasingNoise} to get
 \be
 	H_\mrm{dep}(t)= f(t) \sum_{jk} \ket{j}\bra{k} \bra{j}\sz\ket{k} \eul{i\Delta_{jk}t}.
 \ee

Following closely Ref.~\cite{PhysRevA.79.013819}, we express $f(t)$ in terms of its Fourier decomposition
 \be
 	f(t)=\int_{-\infty}^\infty \diff\omega\;f(\omega)\eul{i\omega t},
 \ee
to obtain
 \be
	 H_\mrm{dep}(t)= \sum_{jk} \sz^{jk} \ket{j}\bra{k}f_{-\Delta_{jk}}(t),
 \ee
where we have defined
 \be
	f_{\Delta_{jk}}(t)=\int_{\Delta_{jk}-B_{jk}}^{\Delta_{jk}+B_{jk}}\diff\omega\;f(\omega)\eul{i(\omega-\Delta_{jk})t}.
 \ee
In writing this expression, we have considered that the main contribution to dephasing comes from a small frequency interval $2B_{jk}$ around $\Delta_{jk}$. For this approximation to be valid, we must have $B_{jk}\ll\Delta_{jk}$. Using the Wiener-Khintchin theorem~\cite{gardiner}
\be
	E[f(\omega)f(-\omega')]=\delta(\omega-\omega')S_f(\omega),
\ee
where $E[x]$ is the classical mean value of $x$ and $S_f(\omega)$ the spectral density of $f(t)$, we then write
\be
	f(\omega)=\sqrt{S_f(\omega)}\xi(\omega),
\ee
with $\xi(v)$ such that $E[\xi(\omega)]=0$ and $E[\xi(\omega)\xi(\omega')]=\delta(\omega-\omega')$, i.e. white noise. We now take $S_f(\omega)$ to be approximately constant over each individual $B_{jk}$ and consider that these bands do not overlap, allowing us to write
\be
	f_{\Delta_{jk}}(t)=\sqrt{S_f(\Delta_{jk})}\int_{-B_{jk}}^{B_{jk}}\diff\omega\;\xi(\omega+\Delta_{jk})\eul{i\omega t}.
\ee
Assuming the dephasing timescale to be much slower than $1/B_{jk}$, we can take $B_{jk}\rightarrow\infty$ and get
\be
	f_{\Delta_{jk}}(t)=\sqrt{S_f(\Delta_{jk})}\xi_{\Delta_{jk}}(t),
\ee
finally leading to
\be
	H_\mrm{dep}(t)=\sum_{jk} \sz^{jk} \ket{j}\bra{k} \xi_{-\Delta_{jk}}(t) \sqrt{S_f(-\Delta_{jk})}.
\ee
If the transition frequencies $\Delta_{jk}$ are well-separated, we can treat each term of the above summation as an independent noise. This last form for $\Hbar_\mrm{dep}(t)$ then yields the following terms in the master equation
\be	\label{eqn-dephasing}
	\sum_{\substack{jk\\j\neq k}} \frac12\gamma_\phi(-\Delta_{jk})|\sz^{jk}|^2 \mathcal{D}\left[\ket{j}\bra{k}\right]+ \frac12\gamma_\phi(0) \mathcal{D}\left[\sum_{j}\Phi_{jj}\ket{j}\bra{j}\right],
\ee
with $\gamma_\phi(-\Delta_{jk})=2 S_f(-\Delta_{jk})$.

\subsection{Quantum model \label{sec-quantum-dephasing}}

To model pure dephasing in a quantum way, we introduce a quantum bath~\cite{carmichael}
\be
	H_\mrm{B} = \sum_j \nu_j b_j^\dagger b_j.
\ee
The interaction of the system with this bath is given by
\be
	H_\mrm{SB}=\sum_{jk}\alpha_{jk}b_j^\dagger b_k\spl\sm,
\ee
where $b_j$ is the ladder operator for bath mode $j$, with frequency $\nu_j$, and $\alpha_{jk}$ is a coupling constant. This interaction corresponds to the transfer of an energy quantum from one bath mode to the other through virtual excitation of the qubit.
We now move to the dressed interaction picture with respect to $H_\mrm S+H_\mrm{B}$
\be
	H_\mrm{I}(t)=\sum_{jk}\alpha_{jk} b^\dag_jb_k\eul{i(\nu_j-\nu_k)t}\eul{iH_\mrm{S}t}\sigma_-\sigma_+\eul{-iH_\mrm{S}t}.
\ee
Using the closure relation of the system, we get
\be
	H_\mrm{I}(t)=\sum_{jkmn}\alpha_{jk} b^\dag_jb_k\eul{i(\nu_j-\nu_k)t}Z_{mn}\ket{m}\bra{n}\eul{i\Delta_{mn}t},
\ee
where we have defined the parity-preserving overlap matrix
\be
	Z_{mn}=\bra{m}{\sigma}_+{\sigma}_-\ket{n}.
\ee

In Appendix \ref{sec-master-relax}, to obtain Eq.~(\ref{eqn-relaxation}) for the coupling to the $X$ and $\sx$ baths, we exploited the fact that all bath operators interacting with the system had zero mean value in thermal equilibrium~\cite{carmichael}. This is \emph{not} the case with the above interaction Hamiltonian, since terms for which $j=k$ have a non-zero thermal mean value. To solve this problem, we include these terms in the system part of the total Hamiltonian, defining an effective shifted Hamiltonian
\be
	H_\mrm{S}'=H_\mrm{S}+\sum_{jmn}\alpha_{jj}b^\dag_jb_jz_{mn}(t),
\ee
where
\be
	z_{mn}(t)=Z_{mn}\ket{m}\bra{n}\eul{i\Delta_{mn}t}.
\ee
Assuming the bath is in thermal equilibrium, we get
\be
	H_\mrm{S}'=H_\mrm{S}+\sum_{jmn}\alpha_{jj}\overline{n}_j(T) z_{mn}(t).
\ee
We can now write the interaction Hamiltonian as
\be
	H_\mrm{I}(t)=B(t)s(t),
\ee
with
\begin{align}
	B(t)&=\sum_{j,k\neq j}\alpha_{jk}b^\dag_jb_k\eul{i(\nu_j-\nu_k)t}\\
	s(t)&=\sum_{mn}z_{mn}(t).
\end{align}
This allows to use the Born master equation
\begin{align}
	&\dot{\rho}_I(t)=-\int_0^t\diff t'\label{eqn-Born-2}\\
		&\times \left\{\left[s(t)s(t')\rho_I(t')-s(t')\rho_I(t')s(t)\right]\mean{B(t)B(t')}_\beta\right.\notag\\
		&\left.+\left[\rho_I(t')s(t')s(t)-s(t)\rho_I(t')s(t')\right]\mean{B(t')B(t)}_\beta\right\},	\notag
\end{align}
where $\beta$ indicates that the mean value is taken in a thermal state. The above correlators take the form
\begin{align}
	\mean{B(t)B(t')}&=\sum_{j,k\neq j}\alpha_{jk}^2\overline{n}_j(T)\left(1+\overline{n}_k(T)\right)\eul{i(\nu_j-\nu_k)\tau}\notag\\
	\mean{B(t')B(t)}&=\sum_{j,k\neq j}\alpha_{jk}^2\overline{n}_j(T)\left(1+\overline{n}_k(T)\right)\eul{-i(\nu_j-\nu_k)\tau},
\end{align}
where $\tau=t-t'$ and where we have taken the system-bath coupling constant to be real and symmetric under the exchange of modes $j$ and $k$. The Born master equation becomes
\begin{align}
	&\dot{\rho}_I(t)=-\sum_{mnm'n'}\notag\\
		& \times\big\{\left[z_{mn}(t)z_{m'n'}(t)\rho_I(t-\tau)-z_{m'n'}(t)\rho_I(t-\tau)z_{mn}(t)\right]\notag\\
		& \times\int_0^t\diff\tau\;\eul{-i\Delta_{m'n'}\tau}\mean{B(t)B(t-\tau)}\notag\\
	&+\left[\rho_I(t-\tau)z_{m'n'}(t)z_{mn}(t)-z_{mn}(t)\rho_I(t-\tau)z_{m'n'}(t)\right]\notag\\
	& \times\int_0^t\diff\tau\;\eul{-i\Delta_{m'n'}\tau}\mean{B(t-\tau)B(t)}\big\},\label{eqn-markov}
\end{align}
Replacing $\rho_I(t-\tau)$ by $\rho_I(t)$ and extending the upper boundary of the integrals over time to infinity, i.e. doing the Markov approximation, we get
\begin{align}
	\int_0^t\diff\tau\;\eul{-i\Delta_{m'n'}\tau}\mean{B(t)B(t-\tau)}&\simeq\frac12\gamma_{m'n'}(T)-iL_{m'n'}\notag\\
	\int_0^t\diff\tau\;\eul{-i\Delta_{m'n'}\tau}\mean{B(t-\tau)B(t)}&\simeq\frac12\gamma_{m'n'}\,\! '(T)-iL_{m'n'}\,\! ',
\end{align}
with
\begin{align}
	\gamma_{mn}=2\pi\int_0^\infty\diff\nu\;&\alpha^2(\nu,\nu+\Delta_{mn})d(\nu)d(\nu+\Delta_{mn})\times\notag\\
	&\overline{n}(\nu,T)\left(1+\overline{n}(\nu+\Delta_{mn},T)\right),
\end{align}
and
\begin{align}
	L_{mn}=P\int_0^\infty\diff\nu\diff\nu'&\;\frac{\alpha^2(\nu,\nu')d(\nu)d(\nu')}{\nu'-\nu-\Delta_{mn}}\notag\\
	&\;\;\;\;\times\overline{n}(\nu,T)(1+\overline{n}(\nu',T)),
\end{align}

As in Appendix~\ref{sec-master-relax}, we assume that all relevant transitions in the system have different frequencies. This allows to drop fast rotating terms. Conditions in which the present approach might be inaccurate are explained in Appendix~\ref{sec:conditions}.

Knowing that
\begin{align}
	Z_{mn}=\frac{\delta_{mn}+\sz^{mn}}{2}\label{id-Zp},
\end{align}
we obtain the following master equation in the Schr\"odinger picture
\begin{align}
	\dot{\rho}(t)=&-i\commut{H_\mrm{S}''}{\rho(t)}+\frac{\gamma_\phi(0)}{2}\mathcal{D}\left[\sum_{m}\sz^{mm}\proj{m}\right]\rho(t)\notag\\
		&+\sum_{m,n\neq m}\frac{\gamma_\phi(\Delta_{nm})}{2}|\sz^{mn}|^2\mathcal{D}[\ket{m}\bra{n}]\rho(t),\label{eqn-master}
\end{align}
with $\gamma_{\phi}(\Delta_{nm})=\gamma_{nm}/2$, and where we have defined the Lamb-shifted Hamiltonian
\begin{align}
	H_S''=&H_S'+\sum_{mn}|Z_{mn}|^2\;L_{mn}\proj{n}.
\end{align}
Equation~(\ref{eqn-master}) is exactly the master equation found for a classical bath if we neglect Lamb shifts, which can be done at low temperature and system-bath coupling.

Finally, since the above master equation has been obtained for a bath in thermal equilibrium, the rates must obey detailed balance~\cite{clerk2010introduction}
\be
	\gamma_\phi(-\omega)=\exp\left(-\frac{\omega}{k_B T}\right)\gamma_\phi(\omega).
\ee

\section{Conditions under which the master equation developed here is applicable \label{sec:conditions}}

Here, we discuss conditions under which all relevant transitions have different frequencies and the above master equation can safely be applied. As stated in Section~\ref{subsec:dispersive}, in the dispersive regime, if $\zeta\sim\kappa$, resonator transitions overlap. On the other hand, if the ratio $g/\Delta$ is large enough to have $\zeta\gg\kappa$, this degeneracy is lifted, at least for low excitation numbers. Indeed, for high excitation numbers, some transitions might accidentally have the same frequency. 

\begin{figure}
	\begin{center}
		\includegraphics[scale=0.7]{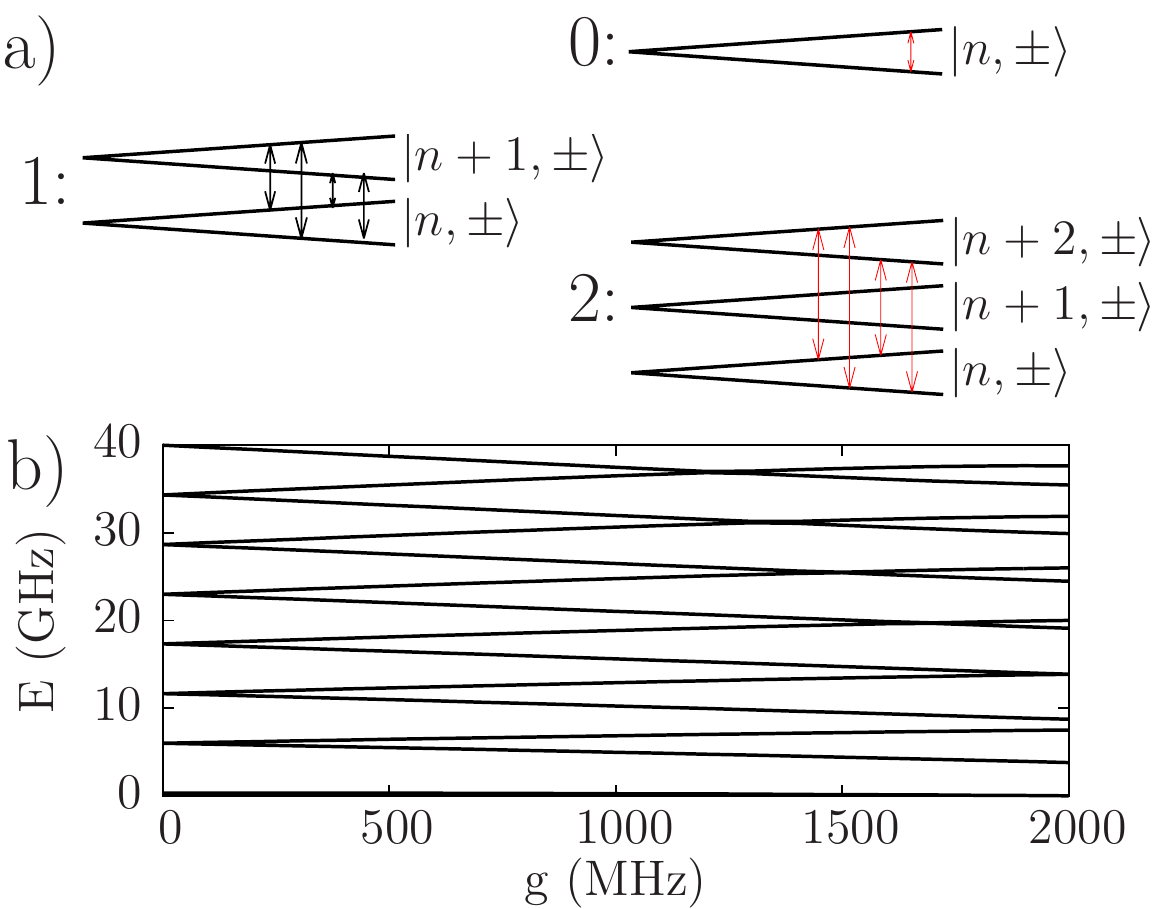}
		\caption{a) Types of transitions allowed in the Bloch-Siegert regime ($g\ll\Sigma$). Red: transitions driven by the (even) $\sz$ bath. Black: transitions induced by the (odd) $X$ and $\sx$ baths. b) Exact energy levels of the Rabi Hamiltonian with increasing coupling strength, obtained numerically. Crossings between levels in the spectrum lead to pairs of transitions with equal frequency. However, since in each of these pairs, one transition is even and the other is odd, they belong to different baths and these overlaps are not relevant for the master equation. Parameters are $\omega_r/2\pi=\omega_a/2\pi=6$ GHz. }
		\label{fig:critical:transitions}
	\end{center}
\end{figure}

We now define a critical excitation number $\tilde n_\mrm{crit}$ under which all transitions occur at different frequencies and thus can be treated independently, given a sufficient ratio $g/\Delta$. We limit ourselves to the Bloch-Siegert regime, under which $g\ll\Sigma$ and the counter-rotating terms are treated in a perturbative way. In this case, the energy levels are
\be
	E_{n,\pm}\simeq n\omega_r\pm\frac12\sqrt{(\Delta_n^\mrm{BS})^2+4g^2n},
\ee
and thus display a nonlinearity scaling in $\sqrt n$. In addition, and as illustrated in Fig.~\ref{fig:critical:transitions}, the bath operators only couple  states in the same doublet, or one or two doublets away from each other. Moreover, as explained in Sec.~\ref{sec-model}, parity selection rules apply so $X$ and $\sx$ baths can drive only transitions between adjacent doublets (Type 1), while $\sz$ noise can induce transitions inside a doublet (Type 0) or between second-nearest-neighbour doublets (Type 2). This allows us to find a distinct $\tilde n_\mrm{crit}$ for individual baths by looking at every possible combination of transitions and finding when some possibly overlap.

We first consider transitions driven by $X$ and $\sx$ baths, which are of Type 1. At low $n$, these essentially occur at distinct frequencies because of the $\sqrt n$ nonlinearity. However, as $n$ increases, $\sqrt n$ becomes comparable to $\sqrt{n+1}$ and these transitions become closer. When the frequency difference between two transitions becomes of the order of $n \kappa$, their typical linewidth, our model breaks down. We get an order of magnitude estimate of this critical $n$ with the condition $(E_{n+1,+}-E_{n,+})-(E_{n+1,-}-E_{n,-})\sim n \kappa$. Dropping terms of order higher than $g^2$ and assuming $g\gg\kappa$ and $g\gtrsim|\Delta|$, i.e. such that the system is \emph{out} of the dispersive regime, we get
\be
	\tilde{n}_\mrm{crit}^{(1)}\simeq\left[\frac{g}{\kappa}\left(1+\frac{\Delta}{2\Sigma}\right)\right]^{2/3}.
\ee
Typically, this means we can have hundreds of excitations before the nonlinearity disappears and the model breaks down for odd baths. As a result, this limitation is not relevant in practice.

We now turn to $\sz$-driven transitions. Type 2 transitions will start to overlap in similar conditions as above, but with higher $n$, since they involve more widely separated energy levels. Thus, these transitions do not set $\tilde n_{crit}$ . Because of the $\sqrt n$ dependence of the nonlinearity, all Type 0 transitions have different frequencies. However, Type 0 and 2 transitions can overlap. These overlaps depend on parameters $g$, $\omega_r$, $\omega_a$, and $n$ in a nontrivial way, but a critical $n$ for which this starts to be possible can be established with the condition $E_{n+1,-}-E_{n-1,+}=E_{n,+}-E_{n,-}$, which leads to a complicated expression for $\tilde n_\mrm{crit}$. Yet, we can get an estimate for this critical number with the criterion $E_{n,+}-E_{n,-}\sim\omega_r$, which leads to
\be
	\widetilde n_\mrm{crit}^{(0-2)}=\frac{\omega_r^2-\Delta^2}{4(g^2+\mu\Delta)}.
\ee
In resonance with $\omega_a/2\pi=\omega_r/2\pi=6$ GHz and for $g/2\pi=1$ GHz, $\widetilde n_\mrm{crit}^{(0-2)}=9$, enough to accurately describe many experiments such as spectroscopy. 

Finally, we emphasize that over $\widetilde n_\mrm{crit}^{(0-2)}$, we can only say that \emph{some} pairs of transitions \emph{might} overlap. If the involved levels do not play a leading role in the dynamics of the system under study, the master equation presented here should still give sensible results in practice.

\section{Transition matrix elements \label{sec-elements}}

In this Appendix, we evaluate the overlap matrix element between eigenstates $\ket{j}$ and $\ket{k}$ of the Rabi Hamiltonian for an arbitrary operator $O$. This is done using the perturbation theory presented in Section~\ref{sec-hamiltonians} to second order in $g$ such that
\begin{align}
	O^{jk}\simeq\bra{j}U^\dagger OU\ket{k},
\end{align}
where $\ket{j}$ is the eigenstate in the Bloch-Siegert basis, Eq.~\eq{eq:BS_Basis_tranformation}. Using Eqs.~(\ref{eqn-nplus}) and~(\ref{eqn-nminus}) with the unitary transformation Eq.~(\ref{eqn-unitary}), we calculate the transition matrix elements Eqs.~(\ref{eqn-sz}), (\ref{eqn-X}) and (\ref{eqn-sx}). For $O=X$, we obtain
{\allowdisplaybreaks\begin{align}
	X^{g0;1-}&=(1+l)\sin\theta_1-l\cos\theta_1\\
	X^{g0;1+}&=(1+l)\cos\theta_1+l\sin\theta_1\notag\\
	X^{n+;n+1,+}&=\left[\sqrt n (1-l)\sin\theta_n+l\cos\theta_n\right]\sin\theta_{n+1}\notag\\
					    &\;\;\;\;+\sqrt{n+1}(1+l)\cos\theta_n\cos\theta_{n+1}\notag\\
	X^{n+;n+1,-}&=-\left[\sqrt n (1-l)\sin\theta_n+l\cos\theta_n\right]\cos\theta_{n+1}\notag\\
					    &\;\;\;\;+\sqrt{n+1}(1+l)\cos\theta_n\sin\theta_{n+1}\notag\\
	X^{n-;n+1,+}&=\left[-\sqrt n (1-l)\cos\theta_n+l\sin\theta_n\right]\sin\theta_{n+1}\notag\\
					    &\;\;\;\;+\sqrt{n+1}(1+l)\sin\theta_n\cos\theta_{n+1}\notag\\
	X^{n+;n+1,-}&=-\left[-\sqrt n (1-l)\cos\theta_n+l\sin\theta_n\right]\cos\theta_{n+1}\notag\\
					    &\;\;\;\;+\sqrt{n+1}(1+l)\sin\theta_n\sin\theta_{n+1}\notag,
\end{align}
where $l=2\xi+l^2/2$; $\xi = g\Lambda/2\omega_r$ is defined below Eq.~(\ref{eqn-unitary}) We note that $X_{ij}=X_{ji}$. All other matrix elements are zero to second order. Similarly, for $O = \sigma_x$ we find
{\allowdisplaybreaks\begin{align}
	\sx^{g0;1-}&=r_0^2\cos\theta_1-s_0\sin\theta_1\\
	\sx^{g0;1+}&=-r_0^2\sin\theta_1-s_0\cos\theta_1\notag\\
	\sx^{n+;n+1,+}&=-\left[r_n^2\sin\theta_{n+1}+s_{n+1}\cos\theta_{n+1}\right]\cos\theta_n\notag\\
		&\;\;\;\;+\left[s_n\sin\theta_{n+1}+t_n\cos\theta_{n+1}\right]\sin\theta_n\notag\\
	\sx^{n+;n+1,-}&=-\left[-r_n^2\cos\theta_{n+1}+s_{n+1}\sin\theta_{n+1}\right]\cos\theta_n\notag\\
		&\;\;\;\;+\left[-s_n\cos\theta_{n+1}+t_n\sin\theta_{n+1}\right]\sin\theta_n\notag\\
	\sx^{n-;n+1,+}&=-\left[r_n^2\sin\theta_{n+1}+s_{n+1}\cos\theta_{n+1}\right]\sin\theta_n\notag\\
		&\;\;\;\;-\left[s_n\sin\theta_{n+1}+t_n\cos\theta_{n+1}\right]\cos\theta_n\notag\\
	\sx^{n-;n+1,-}&=-\left[-r_n^2\cos\theta_{n+1}+s_{n+1}\sin\theta_{n+1}\right]\sin\theta_n\notag\\
		&\;\;\;\;-\left[-s_n\cos\theta_{n+1}+t_n\sin\theta_{n+1}\right]\cos\theta_n,\notag
\end{align}
with $r_n^2=1-\Lambda^2(n+1/2)$, $s_n=\Lambda\sqrt{n}$, and $t_n=2\xi\sqrt{n(n+1)}$. Finally, $O = \sigma_z$ yields
{\allowdisplaybreaks\begin{align}
	\sz^{g0;g0}&=2\Lambda^2-1\label{eqn:elem:sz}\\
	\sz^{g0;2+}&=2\Lambda\sin\theta_2\notag\\
	\sz^{g0;2-}&=-2\Lambda\cos\theta_2\notag\\
	\sz^{n+;n+}&=\left[2\Lambda^2(n-1)-1\right]\cos(2\theta_n)+4\Lambda^2\cos^2\theta_n\notag\\
	\sz^{n+;n-}&=2(2\Lambda^2 n-1)\sin\theta_n\cos\theta_n\notag\\
	\sz^{n-;n-}&=-\left[2\Lambda^2(n-1)-1\right]\cos(2\theta_n)+4\Lambda^2\sin^2\theta_n\notag\\
	\sz^{n+;n+2,+}&=2\Lambda\sqrt{n+1}\cos\theta_n\sin\theta_{n+2}\notag\\
	\sz^{n+;n+2,-}&=-2\Lambda\sqrt{n+1}\cos\theta_n\cos\theta_{n+2}\notag\\
	\sz^{n-;n+2,+}&=2\Lambda\sqrt{n+1}\sin\theta_n\sin\theta_{n+2}\notag\\
	\sz^{n-;n+2,-}&=-2\Lambda\sqrt{n+1}\sin\theta_n\cos\theta_{n+2}.\notag
\end{align}

\section{Photon creation rate under white $\sigma_z$ noise \label{sec-rate}}

In this section, we derive the photon creation rate Eq.~(\ref{eqn-rate}) caused by white noise fluctuations in the qubit transition frequency. To simplify the discussion, we consider only transitions to $\ket{\widetilde{2,\pm}}$, the first accessible doublet.

The photon creation rate is given by
\be
	\beta=\drvst\mean{\ad a}=\Tr\left[\dot\rho(t)\ad a\right].
\ee
We take the initial state to be $\ket{\widetilde{g0}}$. To obtain a constant rate, we limit ourselves to very small times $t$, such that $\beta\simeq\beta(0)$. As illustrated in Fig.~\ref{fig-decoherence}, this will not cause any problem for white noise, for which $\mean{\ad a}$ is found numerically to increase linearly at all times. Since $\dot\rho(0)=-i\commut{H_R}{\rho(0)}+\mathcal{L}_\mrm{dr}\rho(0)$ and $\rho(0)$ commutes with $H_R$, we obtain
\be
	\beta\simeq\Tr\left[ \ad a \mathcal{L}_\mrm{dr}\ket{\widetilde{g0}}\bra{\widetilde{g0}}\right].
\ee
As shown in Eq.~\eq{eqn-lindbladian}, $\mathcal{L}_\mrm{dr}$ has a component responsible for pure dephasing and another that induces transitions. Since $\rho(0)$ is an eigenstate, the dephasing term cancels out. We thus get, after moving to the Bloch-Siegert basis
\begin{align}
	\beta\simeq&-\left(\Gamma\phi^{2-,g0}+\Gamma_\phi^{2+,g0}\right)\bra{g0}(\ad a)^\mrm{BS}\ket{g0}\label{eqn:beta:elements}\\
			  &\!\!\!+\Gamma_\phi^{2-,g0}\bra{2-}(\ad a)^\mrm{BS}\ket{2-}
			    +\Gamma_\phi^{2+,g0}\bra{2+}(\ad a)^\mrm{BS}\ket{2+}.\notag
\end{align}
With
\begin{align}
	(\ad a)^\mrm{BS}=&\;\ad a-\Lambda(a\sm+\ad\spl)-2\xi(a^2+\ad\,\!^2)\sz\notag\\
				&-\Lambda^2\left(\ad a+\frac12\right)\sz+\frac12\Lambda^2,
\end{align}
and using Eq.~\eq{eq:GammaDressedDephasing} for the transition rates as well as Eq.~\eq{eqn:elem:sz} for the corresponding transition matrix elements, we obtain
\be
	\beta\simeq2\Lambda^2\left[T_{2-}(\theta_2)\gamma_\phi(-\omega_{2-})+T_{2+}(\theta_2)\gamma_\phi(-\omega_{2+})\right]\label{eqn:beta},
\ee
where
\begin{align}
	T_{2-}&=(1+\sin^2\theta_2)\cos^2\theta_2\\
	T_{2+}&=(1+\cos^2\theta_2)\sin^2\theta_2,
\end{align}
and $\omega_{2\pm}=E_{2\pm}-E_{g0}$.

Eq.~\eq{eqn:beta} clearly shows that the spectrum at large negative frequencies must be important in the $\sz$ bath for the photon generation rate to be significant. However, in our model, this bath respects detailed balance. Indeed, $\gamma_\phi(-\omega)=\exp(-\omega/k_BT)\gamma_\phi(\omega)$ such that $\gamma_\phi(-\omega)\rightarrow0$ for $\omega\gg k_BT$, meaning that these contributions should be very small for low temperatures.

Finally, when $\gamma(-\omega_{2-})=\gamma(-\omega_{2+})\equiv\gamma_\phi$, which is the case for white noise, Eq.~\eq{eqn:beta} reduces to the simpler Eq.~\eq{eqn-rate}.

\section{Vacuum Rabi splitting \label{sec-rabi}}

As outlined in Section~\ref{sec-asymmetry}, here we calculate $\mean{a}_s$ under weak cavity driving. We will assume that dephasing noise at high negative frequencies is weak, such that transitions from the ground state to the $|\widetilde{2, \pm}\rangle$ doublet are negligible as shown in Appendix~\ref{sec-rate}. Together with the weak driving assumption, this means that only the first three levels of the system are relevant.

For simplicity, we first move to the Bloch-Siegert basis defined by Eq.~\eq{eqn-unitary} to get
\be
	H_\mrm{drvn}^\mrm{BS}(t)= U^\dagger H(t) U = H_\mrm{BS}+\epsilon\,a^\mrm{BS}\eul{i\nu t}+ \mathrm{h.c.}
\ee
It is also useful to move to a rotating frame with
\be
	V(t) = \eul{-i \omega_d [(\ad a)^\mrm{BS}+  \sz^\mrm{BS}/2]t},
\ee
to obtain the time-independent Hamiltonian
\be	\label{eqn-Hdrvn}
	H_\mrm{drvn}^\mrm{BS}=\Delta_r^\mrm{BS}\ad a +\frac{\Delta_a^\mrm{BS}}{2}\sz+gI_++\epsilon(a^\mrm{BS}+\ad\,\!^\mrm{BS}),
\ee
with
\begin{eqnarray}
	\Delta_r^\mrm{BS}=\omega_r-\omega_d-\mu\;\;\;\;\;\;\;	&	\Delta_a^\mrm{BS}=\omega_a-\omega_d+\mu.
\end{eqnarray}

The Heisenberg equation of motion for an arbitrary operator $\hat O$ is
\be\label{eqn:master:O}
	\drvst\mean{\hat O}=i\mean{\commut{H_\mrm{drvn}}{\hat O}}+\mean{\mathcal{L}_\mrm{dr}ï\hat O},
\ee
where in the subspace $\{\ket{\widetilde{g0}}, \ket{\widetilde{1-}},\ket{\widetilde{1+}}\}$
\begin{align}
	\mathcal{L}_\mrm{dr}^O\cdot&= \sum_{\sigma=\pm}\left(\Gamma_\kappa^{g0,1\sigma}+\Gamma_\gamma^{g0,1\sigma}\right)\mathcal{D}_O\left[\ket{\widetilde{g0}}\bra{\widetilde{1\sigma}}\right]\cdot\notag\\
	&+\Gamma_\phi^{1-,1+}\mathcal{D}_O\left[\ket{\widetilde{1-}}\bra{\widetilde{1+}}\right]\cdot
		+\Gamma_\phi^{1+,1-}\mathcal{D}_O\left[\ket{\widetilde{1+}}\bra{\widetilde{1-}}\right]\cdot\notag\\
	&+\mathcal{D}_O\left[\Phi^{g0,g0}\ket{\widetilde{g0}}\bra{\widetilde{g0}}+\Phi^{1-,1-}\ket{\widetilde{1-}}\bra{\widetilde{1-}}\right.\notag\\
	&\;\;\;\;\;\;\;\;\;\;\;\;\;\;\;\;\;\;\;\;\;\;\;\;\;\;\;\;\;\;\;\;\;\;\;\;\;+\left.\Phi^{1+,1+}\ket{\widetilde{1+}}\bra{\widetilde{1+}}\right]\cdot,
\end{align}
where $\mathcal{D}_O[\hat Q]\hat O=(2\hat Q^\dagger\hat O \hat Q - \hat Q^\dagger\hat Q \hat O - \hat O \hat Q^\dagger \hat Q)/2$. Rates are defined in Section~\ref{sec-model}.

We are interested in obtaining the mean value of $a$ and $\sm$. Since mean values does not depend on the frame, we will simplify calculations by working in the Bloch-Siegert picture. To do so, we first calculate the effect of the dissipators on $\aBS$ and $\smbs$, knowing that
\be
	\mathcal{D}_O[\hat Q^\mrm{BS}]\hat O^\mrm{BS}=\left(\mathcal{D}_O[\hat Q]\hat O\right)^\mrm{BS}.
\ee
This means that we can treat the states and operators in the effective Jaynes-Cummings Hamiltonian basis and then use the unitary $U$ defined in Eq.~\eq{eqn-unitary} to move back to the Bloch-Siegert frame, which takes the non-RWA terms in consideration. In the three-level approximation, we have
\begin{align}
	a&=\;\;\;\cos\theta_1\ket{g0}\bra{1+}+\sin\theta_1\ket{g0}\bra{1-}\label{eqn-a}\\
	\sm&=-\sin\theta_1\ket{g0}\bra{1+}+\cos\theta_1\ket{g0}\bra{1-}\label{eqn-sm},
\end{align}
resulting in the dissipators
\begin{align*}
	\mathcal{D}_O[\ket{g0}\bra{1+}]a&=-\frac12\cos\theta_1(a\cos\theta_1-\sm\sin\theta_1)\\
	\mathcal{D}_O[\ket{g0}\bra{1-}]a&=-\frac12\sin\theta_1(a\sin\theta_1+\sm\cos\theta_1)\\
	\mathcal{D}_O[\ket{1-}\bra{1+}]a&=-\frac12\cos\theta_1(a\cos\theta_1-\sm\sin\theta_1)\\
	\mathcal{D}_O[\ket{1+}\bra{1-}]a&=-\frac12\sin\theta_1(a\sin\theta_1+\sm\cos\theta_1)\\
	\mathcal{D}_O[\ket{g0}\bra{1+}]\sm&=+\frac12\sin\theta_1(a\cos\theta_1-\sm\sin\theta_1)\\
	\mathcal{D}_O[\ket{g0}\bra{1-}]\sm&=-\frac12\cos\theta_1(a\sin\theta_1+\sm\cos\theta_1)\\
	\mathcal{D}_O[\ket{1-}\bra{1+}]\sm&=+\frac12\sin\theta_1(a\cos\theta_1-\sm\sin\theta_1)\\
	\mathcal{D}_O[\ket{1+}\bra{1-}]\sm&=-\frac12\cos\theta_1(a\sin\theta_1+\sm\cos\theta_1).
\end{align*}
Proceeding in a similar way for dissipators involved in pure dephasing yields
\begin{align}
	\mathcal{D}_O^\mrm{deph}a=&-\frac12\left(\gamma_\phi^+\cos^2\theta_1+\gamma_\phi^-\sin^2\theta_1\right)a\notag\\
			&-\frac12\sin\theta_1\cos\theta_1\left(\gamma_\phi^--\gamma_\phi^+\right)\sm\\
	\mathcal{D}_O^\mrm{deph}\sm=&-\frac12\sin\theta_1\cos\theta_1\left(\gamma_\phi^--\gamma_\phi^+\right)a\notag\\
			&-\frac12\left(\gamma_\phi^+\sin^2\theta_1+\gamma_\phi^-\cos^2\theta_1\right)\sm,
\end{align}
where we have defined
\be
	\gamma_\phi^\pm=\frac{\gamma_\phi(0)}{2}\left|\sz^{g0,g0}-\sz^{1\pm,1\pm}\right|^2.
\ee
If we now add the contributions from all dissipators in Eq.~(\ref{eqn:master:O}) to the Heisenberg equation for $a$, we get
\be
	-\Gamma_+(\theta_1)\mean a-\eta(\theta_1)\mean{\sm},
\ee
while for  $\sm$, we obtain
\be
	-\eta(\theta_1)\mean a-\Gamma_-(\theta_1)\mean{\sm}.
\ee
Here, we have defined
\begin{align}
	\Gamma_+(\theta_1)&=\Gamma_1\sin^2\theta_1
						+\Gamma_2\cos^2\theta_1\\
	\Gamma_-(\theta_1)&=\Gamma_1\cos^2\theta_1
						+\Gamma_2\sin^2\theta_1\\
	\eta(\theta_1)&=\left(\Gamma_1-\Gamma_2\right)\sin\theta_1\cos\theta_1,
\end{align}
with the rates
\begin{align}
	\Gamma_1=\frac{\gamma_-+\gamma_\phi^\uparrow+\gamma_\phi^-}{2};\;\;\;\;\;\;
		&\Gamma_2=\frac{\gamma_++\gamma_\phi^\downarrow+\gamma_\phi^+}{2}.
\end{align}
This in turn involves the expressions
\begin{align}
	\gamma_\pm&=\kappa(\Delta_{1\pm,g0})\left|X^{g0,1\pm}\right|^2\notag\\&\;\;\;\;\;\;\:\;\:\;\:\;\:+\gamma(\Delta_{1\pm,g0})\left|\sx^{g0,1\pm}\right|^2,\\
	\gamma_\phi^{\uparrow/\downarrow}&=\frac12\gamma_\phi(\Delta_{1\mp,1\pm})\left|\sz^{1\mp,1\pm}\right|^2.
\end{align}
We now calculate $\commut{H_\mrm{drvn}}{\aBS}$ and $\commut{H_\mrm{drvn}}{\szbs}$.
Neglecting terms that lead to leakage out of the effective Jaynes-Cummings three-level system, we obtain simple forms for $\aBS$ and $\szbs$ in the bare basis
\begin{align}
	a^\mrm{BS}&\approx \left(1+\frac{\Lambda^2}{2}\right)a-\Lambda\spl+2\xi\ad\label{eqn-abs}\\
	\smbs&\approx\left(1+\frac{\Lambda^2}{2}\right)\sm-\Lambda\ad.\label{eqn-smbs}
\end{align}
From these expressions, we easily get
\begin{align}
	\commut{H_\mrm{drvn}}{\aBS}&\simeq-\epsilon-\left(1+\frac{\Lambda^2}{2}\right)\drbs a-\left(1+\frac{\Lambda^2}{2}\right)g\sm\notag\\
		&\!\!\!\!\!+(2\xi\drbs-\mu)\ad+(2\xi g-\Lambda\dqbs)\spl\\
	\commut{H_\mrm{drvn}}{\smbs}&\simeq-\left(1+\frac{\Lambda^2}{2}\right)\dqbs \sm-\left(1+\frac{\Lambda^2}{2}\right)g a\notag\\
			&\;\;\;\;-\Lambda\drbs\ad-\mu\spl.
\end{align}
We now want to express this result in the Bloch-Siegert basis. 
From Eqs.~(\ref{eqn-abs}) and~(\ref{eqn-smbs})
\be
	\sm\simeq\left(1+\frac{\Lambda^2}{2}\right)\smbs+\Lambda\adbs.
\ee
Knowing that $\aBS=a+\mathcal{O}(\Lambda)$, Eq.~(\ref{eqn-abs})
leads to
\be
	a\simeq\left(1-\frac{\Lambda^2}{2}\right)\aBS+\Lambda\spbs-2\xi\adbs.
\ee
This allows to find the appropriate commutators
\begin{align}
	\commut{H_\mrm{drvn}}{\aBS}\simeq&-\epsilon-\drbs \aBS-\left(1+\frac{\Lambda^2}{2}\right)g\sm\notag\\
		&+2(\xi\drbs-\mu)\adbs\notag\\
		&-[\Lambda(\dqbs+\drbs)-2\xi g]\spbs,\\
	\commut{H_\mrm{drvn}}{\smbs}\simeq&-\left[(1+\Lambda^2)\dqbs+\Lambda^2\drbs\right] \smbs-g \aBS\notag\\
			&\!\!\!\!\!\!\!\!\!\!\!\!\!\!\!\!\!\!\!\!\!-\left[\Lambda(\dqbs+\drbs)-2\xi g\right]\adbs-2\mu\spbs.
\end{align}
We can now write equations for the evolution of $\aBS$ and $\smbs$. While Hamiltonian contributions are purely imaginary, those coming from dissipation are purely real. Imaginary terms lead to oscillatory behaviour, while real terms account for excitation and relaxation. Terms in $\aBS$ and $\smbs$ have both real and imaginary components, but terms in $\adbs$ and $\spbs$ only have imaginary contributions. The latter then only contribute through oscillations  in the dynamics. Since we are only interested in the steady-state behavior, we can neglect them. This allows to get the following equations for the steady state if we neglect terms of order higher than $g^2$
\begin{align*}
	\left[i\drbs+\Gamma_+(\theta_1)\right]\mean{\aBS}+\left[ig+\eta(\theta_1)\right]\mean{\smbs}+i\epsilon&=0\\
	\left[ig+\eta(\theta_1)\right]\mean{\aBS}+\left[i\tilde\Delta_a^\mrm{BS}+\Gamma_-(\theta_1)\right]\mean{\smbs}&=0,
\end{align*}
where we have defined $\tilde\Delta_a^\mrm{BS}=(1+\Lambda^2)\dqbs+\Lambda^2\drbs$.
Solving the above set of equations, we finally obtain
\be	\label{eqn-aBS}
	\mean{a}_s=\frac{i\epsilon\,G_q(\theta_1)}
				{G_\eta^2(\theta_1)-G_q(\theta_1)G_r(\theta_1)},
\ee
where
\begin{align}
	G_q(\theta_1)&=\Gamma_-(\theta_1)+i\tilde\Delta_a^\mrm{BS}\\
	G_r(\theta_1)&=\Gamma_+(\theta_1)+i\drbs\\
	G_\eta(\theta_1)&=ig+\eta(\theta_1).
\end{align}

\bibliography{pra}

\end{document}